\documentstyle[preprint,aps,prl]{revtex}

\begin{document}

\draft

\title{Thermodynamics of a trapped Bose-condensed gas} 

\author{ S. Giorgini$^{1,2}$, L. P. Pitaevskii$^{2,3,4}$ and S. Stringari$^{2}$}

\address{$^{1}$ECT$\ast$, Villa Tambosi, Strada delle Tabarelle 286, I-38050 
Villazzano, Italy} 
\address{$^{2}$Dipartimento di Fisica, Universit\`a di Trento, \protect\\
and Istituto Nazionale di Fisica della Materia, I-38050 Povo, Italy}
\address{$^{3}$Department of Physics, Technion, 32000 Haifa, Israel}
\address{$^{4}$Kapitza Institute for Physical Problems, 117454 Moscow, Russia}


\maketitle

\begin{abstract}

{\it We investigate the thermodynamic behaviour of a Bose gas interacting with 
repulsive forces and confined in a harmonic anisotropic trap. We develop the 
formalism of mean field theory for non uniform systems at finite temperature,
based on the generalization of Bogoliubov theory for uniform gases.
By employing the WKB semiclassical approximation for the excited states we 
derive systematic results for the temperature dependence of various 
thermodynamic quantities: condensate fraction, density profiles, thermal 
energy, specific heat and moment of inertia. Our analysis points out important
differences with respect to the thermodynamic behaviour of
uniform Bose gases. This is mainly the consequence of a major role 
played by single particle states at the boundary of the condensate. We find  
that the thermal depletion of the condensate is strongly enhanced by the 
presence of repulsive interactions and that the critical temperature is 
decreased with respect to the predictions of the non-interacting model. 
Our work points out 
an important 
scaling behaviour exhibited by the system in large $N$ limit. Scaling permits 
to express all the relevant thermodynamic quantities in terms of only two 
parameters: the reduced temperature $t=T/T_c^0$ and the ratio between the 
$T=0$ value of the chemical potential and the critical temperature $T_c^0$ 
for Bose-Einstein condensation. Comparisons with first experimental results 
and ab-initio calculations are presented.} 

\end{abstract}


\narrowtext

\section{Introduction}

The experimental realization of Bose-Einstein condensation (BEC) in atomic 
gases \cite{EXP}  
confined in magnetic traps has stimulated a novel interest in the theory
of interacting Bose gases. 
In particular the recent experiments carried out 
at Mit on $^{23}$Na \cite{MIT} and at Jila on $^{87}$Rb \cite{JIL}
have provided first quantitative results for relevant thermodynamic quantities,
such as the temperature dependence of the condensate fraction and the release 
energy.
Based on the pioneering works by 
Bogoliubov \cite{BOG} and Huang, Yang, Luttinger and Lee \cite{HYLL}, 
the theory of 
interacting Bose gases
has so far been mainly focused on 
uniform systems. The possibility of extending it to non-uniform gases, first
explored in the works by Gross, Pitaevskii and Fetter \cite{GP,FET}, and 
stimulated
by the recent experimental activity, is expected to open
new challenging perspectives in many-body physics,
since these systems are characterized
by new length scales and 
symmetries. The thermodynamic behaviour of trapped Bose gases has been already 
the object of several theoretical works, starting from the investigation of 
the ideal Bose gas \cite{BPK}, the development of the Hartree-Fock formalism
\cite{HF}, the inclusion of interaction effects using the local density 
approximation \cite{LD1,LD2}, up to the most recent approaches 
based on self-consistent mean-field theory \cite{GRI,GPS,HZG,SNS} and 
numerical 
simulations \cite{KRA}.   

The purpose of this paper is to provide a systematic discussion of the  
properties of these systems at thermodynamic equilibrium. 
We will always assume that the dilute gas condition,
\begin{equation}
n a^3 << 1 \;\;,
\label{dilute}
\end{equation}
is satisfied. This condition is reached in all the available experiments.
Here $n$ is the atomic density whose typical values range from $10^{11}$
to $10^{15}$ cm$^{-3}$, while the triplet spin s-wave
scattering length $a$ can reach a few tens Angstrom. 
The corresponding value of $n a^3$ is then always extremely small
(10$^{-4}$ - 10$^{-5}$) 
and one would naively expect that the role of interactions is a minor 
effect in such systems, at least for the equilibrium properties. 
This is not the case for these trapped atoms, 
because of the combined effect of 
trapping and Bose-Einstein condensation. To illustrate
this important feature let us consider the simplest case
of $N$ atoms occupying the ground state of an harmonic oscillator  trap.
The ratio between the interaction energy and the quantum oscillator 
energy $\hbar \omega$ behaves as:
\begin{equation}
{E_{2-body}\over N\hbar \omega} \sim N \frac{a}{a_{HO}} \;\;,
\label{ratio}
\end{equation}
where $a_{HO}=(\hbar/m\omega)^{1/2}$ is the harmonic oscillator length.
In the available experiments $a/a_{HO}$ is $\sim 10^{-3}$
while $N$ ranges between $10^3$ and $10^7$ so that the ratio (\ref{ratio}) 
can be very large, revealing that at least the ground state properties 
can never be calculated treating the interaction potential
in a perturbative way.
Differently from what happens in a homogeneous Bose gas where the density
is kept fixed, the repulsion has here the effect of expanding
the condensate wave function. Actually,
for sufficiently large values of $N$, it happens that the $T=0$ equilibrium
properties of these new systems are governed by the competition between the 
2-body force and the external potential, the kinetic energy playing a 
minor role. In this regime the ground state density is
well approximated by the Thomas-Fermi (TF) approximation
\begin{equation}
n({\bf r}) = \frac{m}{4\pi\hbar^2a} (\mu_0 - V_{ext}({\bf r})) \;\;,
\label{thfe}
\end{equation}
if $V_{ext}({\bf r}) \le \mu_0$ and $n({\bf r})=0$ elsewhere.
Here $\mu_0$ is the $T=0$ chemical potential, fixed by the 
normalization condition.
Assuming a harmonic potential $V_{ext} =m(\omega_x^2x^2+\omega_y^2y^2+
\omega_z^2z^2)/2$, the chemical potential
takes the useful expression \cite{OX1,BP}
\begin{equation}
\mu_0 = \frac{\hbar\omega}{2}\left( 15N\frac{a}{a_{HO}}\right)^{2/5} \;\;,
\label{thfe1}
\end{equation} 
where the frequency $\omega$ is 
the geometrical average 
\begin{equation}
\omega=(\omega_x
\omega_y\omega_z)^{1/3}
\label{geomav}
\end{equation}
of the oscillator frequencies and fixes the harmonic oscillator length
\begin{equation}
a_{HO} = \left(\frac{\hbar}{m\omega}\right)^{1/2} \;\;.
\label{aHO}
\end{equation} 
The root mean square (r.m.s.) radius of the condensate takes the form 
\begin{equation}
\sqrt{\langle r^2\rangle}=a_{HO}\left(15N\frac{a}{a_{HO}}\right)^{1/5}
\frac{\sqrt{2\lambda^2+1}}{\sqrt{7}\lambda^{2/3}} \;\;,
\label{meanr}
\end{equation}
where $\lambda$ is the anisotropy parameter of a cylindrically simmetric trap 
$\lambda=\omega_x/\omega_z=\omega_y/\omega_z$. In the presently available 
traps one has typically $a_{HO}=10^{-4}$ cm and $\sqrt{\langle r^2\rangle}$ 
ranges from 2 to 100 $\times 10^{-4}$ cm.
The validity of the TF approximation (\ref{thfe}), (\ref{thfe1}) is based 
on the 
condition $N a/a_{HO} \gg 1$ which also implies $\mu_0/\hbar\omega \gg 1$.
These results have been obtained at $T=0$. They play however a relevant role
also at finite temperature. In particular the value of $\mu_0$ fixes an 
important reference in the scale of temperatures. For $k_BT$ smaller than 
$\mu_0$
the thermodynamic behaviour of the system is in fact affected 
by collective effects (phonon regime in a homogeneous superfluid).
Conversely for higher temperatures 
it is dominated by single-particle
effects. 

It is highly interesting to compare the value of the chemical potential
$\mu_0$ with the one of the critical temperature. Due to the minor role
played by interactions at high $T$, a reliable estimate of the 
critical temperature  
is obtained using the non-interacting model. In this model
the Hamiltonian is given by $H= \sum_i (p^2_i/2m + 
V_{ext}({\bf r}_i))$ and its thermodynamic behaviour can be worked out
exactly. For sufficiently large numbers of atoms this model predicts
the value \cite{BPK,N1} 
\begin{equation}
k_BT_c^0 = \hbar\omega \left( \frac{N}{\zeta(3)} \right)^{1/3} 
\label{Tc}
\end{equation}
for the critical temperature. In the presently 
available traps one has 
typically $\hbar \omega/k_B = 3-5$ nK, while $T_c^0$ ranges from
50 to 300 nK.
Using the TF result (\ref{thfe1})  
the ratio $\eta$ between the chemical potential and the critical temperature
takes the useful form 
\begin{equation}
\eta = {\mu_0 \over k_BT_c^0} = \alpha (N^{1/6}
{a\over a_{HO}})^{2/5}
 \simeq 2.24 (a^3 n_{T=0}({\bf r}=0))^{1/6} \;\;. 
\label{ratio1}
\end{equation}
where $\alpha ={1\over 2}\zeta(3)^{1/3}15^{2/5}\simeq 1.57$ and 
we have used result (\ref{thfe}) for the density. 
This equation reveals an interesting  behaviour of these trapped Bose gases.
In fact although the gas parameter $n a^3$ is always very small, 
its $1/6$-th power can be easily of the order of unity and in 
actual experiments it turns out that $\mu_0$ is  0.3-0.4 $k_BT_c^0$. 
Furthermore the ratio (\ref{ratio1}) depends on  the 
number $N$ of atoms in the trap in an extremely smooth way ($\eta 
\sim N^{1/15}$).

The ratio $\mu_0/k_BT^0_c$ plays a crucial role to characterize
the interplay between interaction and thermal effects. In particular
we will show that the thermodynamic properties of the Bose
condensed gases, for sufficiently large values of $N$, depend on the 
relevant parameters of the system (number $N$ of atoms in the trap,
s-wave scattering length $a$, oscillator length and deformation of the trap)
through
the dimensionless scaling parameters $\eta$.
Note that such a parameter involves a different
combination of the quantities $N$ and $a/a_{HO}$, as compared 
to the ratio (\ref{ratio}) which is instead the natural scaling parameter 
for the ground state properties \cite{OX1,BP,DS}.

In the paper we will investigate the thermodynamic behaviour 
of the trapped Bose gas using a 
mean-field theory based on already available extensions 
of Bogoliubov theory to finite temperature. In particular we will work
in the so called Popov approximation. The details of this approximation
are discussed in Sect. 2. Here we recall some major features. 
At $T=0$ the theory exactly accounts for the Bogoliubov 
spectrum of elementary excitations. At high $T$ 
it approaches the finite-temperature Hartree-Fock theory \cite{HF} 
with which it exactly
coincides  above the critical temperature. 
These two important features (collective and single-particle effects) are 
the main ingredients 
of the Popov approximation, which is then expected to be a good approximation 
in the whole range of temperature except near $T_c$ where mean-field theories 
are known to fail. 
An interesting feature of these trapped gases, 
which distinguishes them from uniform Bose gases, is that single-particle
excitations play an important role also at low temperatures. The reason is 
that these systems exhibit two kinds of low lying excitations: 
on the one hand one has collective modes of the condensate, on the other hand 
one has single-particle excitations near the classical boundary.
In the paper we discuss the relative importance of the corresponding 
contributions to thermodynamics. We find that the single-particle 
contribution is always relevant. In particular, even at low temperatures, 
the thermal energy is governed by the excitation of single-particle states 
differently from what happens in homogeneous Bose systems. 
This fact explains 
why in the trapped gases the Hartree-Fock approximation provides a very good 
description for most thermodynamic quantities in a wide range of temperatures.  

In order to simplify the formalism and to carry out explicit calculations
in a systematic way we will restrict our analysis 
to temperatures such that  
\begin{equation}
k_BT >> \hbar \omega \;\;.
\label{Tcond}   
\end{equation}
This assumption allows us to describe the motion of the thermally excited atoms
in the WKB semi-classical approximation \cite{N2}. In a deformed trap 
condition (\ref{Tcond}) is separately imposed on the three oscillator energies 
$\hbar\omega_x$, $\hbar\omega_y$ and $\hbar\omega_z$.  
Current experimental investigations are carried out at temperatures 
well consistent with the above condition.
The study of the thermodynamic behaviour 
for lower temperatures (of the order of the oscillator 
temperature) requires a quantum description 
of the low lying
elementary modes, beyond the semi-classical approximation.
These modes have been the object of recent extensive experimental
and theoretical work \cite{JIL1,MIT1,TCE,OXF,HZG}.

The final equations resulting from the mean-field approach have the form
of coupled equations for the order parameter and for the density
of the thermally excited atoms. Their numerical solution
will be systematically discussed during the paper and compared 
with the available experimental data as well as with the first results
obtained with Path Integral Monte Carlo simulations with which we find a
remarkable agreement \cite{KRA}. 

The paper is organized as follows. Sect. 2 is devoted to the presentation of
the theoretical
formalism. We discuss the Popov approximation, we obtain formal results for 
the relevant thermodynamic quantities and we derive explicitly the scaling 
behaviour as well as the low $T$ expansion of the thermodynamic functions. 
In Sect. 3 we discuss our results 
for the temperature dependence of  
the condensate fraction, density profile, energy, specific 
heat and moment of inertia.  
Finally in Sect. 4 we make a comparison between
the properties of a trapped  Bose gas and the ones of the uniform gas. 
 
\section{Theory}

\subsection{Self-consistent Popov approximation}

The self-consistent mean-field method described here follows the treatment of 
the inhomogeneous dilute Bose gas developed in Ref. \cite{FET} within  
the Bogoliubov approximation at $T=0$ and its extension to the 
finite-temperature 
Hartree-Fock-Bogoliubov (HFB) approximation recently discussed by Griffin 
in Ref. \cite{GRI}.
The starting point is the grand-canonical Hamiltonian written in terms of the 
creation and annihilation particle field operators $\psi^{\dagger}({\bf r})$ 
and $\psi({\bf r})$ 
\begin{eqnarray}
K  \equiv H-\mu N &=& \int\;d{\bf r} \psi^{\dagger}({\bf r}) 
\left( -\frac{\nabla^2}
{2m} + V_{ext}({\bf r}) - \mu \right) \psi({\bf r}) \nonumber \\
&+& \frac{g}{2} \int\;d{\bf r} \psi^{\dagger}({\bf r})\psi^{\dagger}({\bf r})
\psi({\bf r})\psi({\bf r}) \;\;.
\label{gch}
\end{eqnarray}
In the above equation $V_{ext}({\bf r})=m(\omega_x^2x^2+\omega_y^2y^2
+\omega_z^2z^2)/2$ is the anisotropic harmonic potential which provides the 
confinement of the atoms and $g=4\pi\hbar^2a/m$ is the interaction coupling 
constant fixed by the s-wave scattering length $a$. By separating out the 
condensate part of the particle field operator one can write $\psi({\bf r})$ 
as the sum of the spatially varying condensate wave-function $\Phi({\bf r})$ 
and the operator $\tilde{\psi}({\bf r})$, which acts on the non-condensate 
particles, 
\begin{equation}
\psi({\bf r}) = \Phi({\bf r}) + \tilde{\psi}({\bf r}) \;\;.
\label{parop}
\end{equation}
The wave-function $\Phi$ is formally defined as the statistical average of the 
particle field operator, $\Phi({\bf r}) = \langle\psi({\bf r})\rangle$, and 
represents the order parameter of the system in the condensate phase where 
gauge symmetry 
is broken \cite{HM}. 
Using decomposition (\ref{parop}) the interaction term
in (\ref{gch}) becomes (all quantities depend on ${\bf r}$)
\begin{eqnarray}
\psi^{\dagger}\psi^{\dagger}\psi\psi &=& |\Phi|^4 + 2|\Phi|^2\Phi^{\ast}
\tilde{\psi} + 2|\Phi|^2\Phi\tilde{\psi}^{\dagger} \nonumber \\
&+& 4|\Phi|^2\tilde{\psi}^{\dagger}\tilde{\psi} + \Phi^{\ast 2}\tilde{\psi}
\tilde{\psi} + \Phi^2\tilde{\psi}^{\dagger}\tilde{\psi}^{\dagger} 
\label{inter}  \\
&+& 2\Phi^{\ast}\tilde{\psi}^{\dagger}\tilde{\psi}\tilde{\psi} 
+ 2\Phi\tilde{\psi}^{\dagger}\tilde{\psi}^{\dagger}\tilde{\psi}
+ \tilde{\psi}^{\dagger}\tilde{\psi}^{\dagger}\tilde{\psi}\tilde{\psi}\;\;.
\nonumber
\end{eqnarray}
In the following the cubic and quartic terms in $\tilde{\psi}$, $\tilde{\psi}^
{\dagger}$ in the last line of the above equation will be treated in the 
mean-field approximation. For the cubic terms one gets 
\begin{eqnarray}
\tilde{\psi}^{\dagger}({\bf r})\tilde{\psi}({\bf r})\tilde{\psi}({\bf r}) &=&
2\langle\tilde{\psi}^{\dagger}({\bf r})\tilde{\psi}({\bf r})\rangle\tilde{\psi}
({\bf r}) \equiv
2n_T({\bf r})\tilde{\psi}({\bf r}) \;\;,\nonumber \\
\tilde{\psi}^{\dagger}({\bf r})\tilde{\psi}^{\dagger}({\bf r})\tilde{\psi}
({\bf r}) &=& 
2\langle\tilde{\psi}^{\dagger}({\bf r})\tilde{\psi}({\bf r})\rangle\tilde{\psi}
^{\dagger}({\bf r}) \equiv
2n_T({\bf r})\tilde{\psi}^{\dagger}({\bf r}) \;\;,
\label{cub}
\end{eqnarray}
where $n_T({\bf r})$ is the non-condensate density of particles. 
Whereas the quartic term becomes
\begin{equation} 
\tilde{\psi}^{\dagger}({\bf r})\tilde{\psi}^{\dagger}({\bf r})\tilde{\psi}
({\bf r})\tilde{\psi}({\bf r}) = 
4\langle\tilde{\psi}^{\dagger}({\bf r})\tilde{\psi}({\bf r})\rangle\tilde{\psi}
^{\dagger}({\bf r})\tilde{\psi}({\bf r}) \equiv
4n_T({\bf r})\tilde{\psi}^{\dagger}
({\bf r})\tilde{\psi}({\bf r})  \;\;.
\label{qua}
\end{equation}
To obtain the mean-field factorizations (\ref{cub}) and (\ref{qua}) we have 
neglected the terms proportional to the  
anomalous non-condensate density    
$m_T({\bf r})=\langle\tilde{\psi}({\bf r})\tilde{\psi}
({\bf r})\rangle$, and to its complex conjugate.
This approximation, which is discussed in depth in Ref. 
\cite{GRI}, corresponds to the so called Popov approximation \cite{POP}, and is
expected to be appropriate both at high temperatures, where 
$n_T \gg m_T$, and at low temperatures where the two densities are of the same  
order, but both negligibly small for the very dilute systems we are 
considering here. 
Using decomposition (\ref{parop}) and the mean-field prescriptions 
(\ref{cub}), and (\ref{qua}) the grand-canonical Hamiltonian (\ref{gch}) can
be written as the sum of four terms
\begin{equation}
K = K_0 + K_1 + K_1^{\dagger} + K_2 \;\;.
\label{gch1}
\end{equation}
The first term 
$K_0$ contains only the condensate wave function $\Phi({\bf r})$
\begin{equation}
K_0 = \int\;d{\bf r} \Phi^{\ast}({\bf r}) \left( - \frac{\nabla^2}{2m} 
+ V_{ext}({\bf r}) - \mu + \frac{g}{2} n_0({\bf r}) \right)\Phi({\bf r}) 
\;\;,
\label{K0}
\end{equation}
where we have introduced the condensate density $n_0({\bf r})=|\Phi({\bf r})|^2
$. The 
$K_1$ and $K_1^{\dagger}$ terms are linear in the operators 
$\tilde{\psi}$, $\tilde{\psi}^{\dagger}$
\begin{equation}
K_1 = \int\;d{\bf r} \tilde{\psi}^{\dagger}({\bf r}) \left( - \frac{\nabla^2}
{2m} + V_{ext}({\bf r}) - \mu + g(n_0({\bf r}) + 2n_T({\bf r})) \right)
\Phi({\bf r}) \;\;,
\label{K1}
\end{equation}
and finally $K_2$ is quadratic in $\tilde{\psi}$, $\tilde{\psi}^{\dagger}$
\begin{eqnarray}
K_2 &=& \int\;d{\bf r} \tilde{\psi}^{\dagger}({\bf r}) {\cal L} \tilde{\psi}
({\bf r})  + \frac{g}{2}\int\;d{\bf r} \Phi^2({\bf r}) \tilde{\psi}^{\dagger}
({\bf r}) \tilde{\psi}^{\dagger}({\bf r}) \nonumber \\
&+& \frac{g}{2}\int\;d{\bf r} \Phi^{\ast 2}({\bf r}) \tilde{\psi}({\bf r})
\tilde{\psi}({\bf r}) \;\;.
\label{K2}
\end{eqnarray}
In eq. (\ref{K2}), to make notations simpler, we have introduced the hermitian 
operator
\begin{equation}
{\cal L} = - \frac{\nabla^2}{2m} + V_{ext}({\bf r}) - \mu + 2gn({\bf r}) \;\;,
\label{Lop}
\end{equation}
where 
\begin{equation}
n({\bf r}) = 
n_0({\bf r}) + n_T({\bf r})
\label{totaldensity}
\end{equation}
is the total particle density of the system.
The $K_1$, $K_1^{\dagger}$ terms in the sum (\ref{gch1}) vanish if 
$\Phi({\bf r})$ is the solution 
of the equation for the condensate 
\begin{equation}
\left[ - \frac{\nabla^2}{2m} + V_{ext}({\bf r}) - \mu 
+ g(n_0({\bf r}) + 2n_T({\bf r}))\right] \Phi({\bf r}) = 
\left[ {\cal L} - gn_0({\bf r}) \right] \Phi({\bf r}) = 0 \;\;,
\label{cequ}
\end{equation}
with ${\cal L}$ defined in eq. ({\ref{Lop}), 
whereas the quadratic term $K_2$ can be diagonalized by means of the 
Bogoliubov linear transformation for the particle operators $\tilde{\psi}$, 
$\tilde{\psi}^{\dagger}$
\begin{eqnarray}
\tilde{\psi}({\bf r}) &=& \sum_j ( u_j({\bf r}) \alpha_j + v_j^{\ast}({\bf r})
\alpha_j^{\dagger}) \;\;, \nonumber \\
\tilde{\psi}^{\dagger}({\bf r}) &=& \sum_j (u_j^{\ast}({\bf r})
\alpha_j^{\dagger} + v_j({\bf r})\alpha_j) \;\;.
\label{trans}
\end{eqnarray}
In the above transformation $\alpha_j$, 
and $\alpha_j^{\dagger}$ are quasi-particle annihilation and creation 
operators which satisfy  
the usual Bose commutation relations, whereas the functions 
$u_j({\bf r})$, $v_j({\bf r})$ obey to the following normalization 
condition 
\begin{equation}
\int\;d{\bf r} (u_i^{\ast}({\bf r})u_j({\bf r}) - v_i^{\ast}({\bf r})
v_j({\bf r})) = 
\delta_{ij} \;\;.
\label{ortcon}
\end{equation}
The grand-canonical Hamiltonian 
(\ref{gch}), in the Popov mean-field approximation, takes the form  
\begin{equation}
K_{POPOV} = \int\;d{\bf r} \Phi({\bf r}) \left( - \frac{\nabla^2}{2m} 
+ V_{ext}({\bf r}) - \mu + \frac{g}{2}n_0({\bf r}) \right)\Phi({\bf r})
+ \sum_j \epsilon_j \alpha_j^{\dagger}\alpha_j \;\;,
\label{gch2}
\end{equation}
where the real wave-function $\Phi({\bf r})$ is 
the solution of eq. (\ref{cequ}) 
and the quasi-particle 
energies $\epsilon_j$ are obtained
from the solutions of the coupled Bogoliubov equations 
\begin{eqnarray}
{\cal L}u_j({\bf r}) + gn_0({\bf r})v_j({\bf r}) &=& \epsilon_ju_j({\bf r}) 
\nonumber \\
{\cal L}v_j({\bf r}) + gn_0({\bf r})u_j({\bf r}) &=& - \epsilon_j
v_j({\bf r}) 
\label{bogeqs}
\end{eqnarray}
for the functions
$u_j({\bf r})$ and $v_j({\bf r})$.
In eq. (\ref{gch2}) we have neglected the term $-\sum_j\epsilon_j
\int d{\bf r}|v_j({\bf r})|^2$ which at $T=0$ accounts for the energy of the 
non-condensate particles in the ground state. This is a good approximation 
since, for dilute systems satisfying condition (\ref{dilute}), the
quantum depletion $\sum_j\int d{\bf r}|v_j({\bf r})|^2$ of the condensate is a
minor effect.   
Explicit calculations show that this depletion is always
smaller than one percent of the total number of atoms in the condensate 
\cite{HZG}. 
We have therefore at $T=0$, $\psi({\bf r})=\Phi({\bf r})$ and $n_T({\bf r})=
m_T({\bf r})=0$, 
whereas at finite temperatures we take $n_T$ and $m_T$ proportional to the 
thermal 
density of quasi-particles 
\begin{equation}
n_T=\sum_j(|u_j|^2+|v_j|^2)\langle\alpha_j^{\dagger} 
\alpha_j\rangle \;\;,
\label{nT}
\end{equation}
and
\begin{equation}
m_T=2\sum_j u_jv_j^{\ast}\langle\alpha_j^{\dagger}\alpha_j\rangle \;\;.
\label{mT}
\end{equation}
  
Equations (\ref{bogeqs}) have the typical  form of the equations
of the random phase approximation and have been recently solved in 
Ref. \cite{OXF} at $T=0$; 
the resulting energies $\epsilon_j$ 
give the excitation frequecies of the collective modes of the system.
Very recently, first solutions 
have become available also at finite temperature \cite{HZG}.
The coupled equations (\ref{bogeqs}) can be solved very easily if one uses the 
semiclassical \cite{GPS} or local density 
approximation
\cite{LD1,LD2}.
We write
\begin{equation}
u_j({\bf r}) = u({\bf p},{\bf r})e^{i\varphi({\bf r})} \;\;\;\;\;\;
v_j({\bf r}) = v({\bf p},{\bf r})e^{i\varphi({\bf r})} \;\;, 
\label{sclas}
\end{equation}
where the impulse of the elementary excitation is fixed 
by the gradient 
of the phase
${\bf p}=\hbar\nabla\varphi$ and satisfies 
the condition $p\gg\hbar/a_{HO}$. 
We also assume that $u({\bf p},{\bf r})$, $v({\bf p},{\bf r})$ (and hence 
$n_T$)  
are smooth functions of ${\bf r}$ on length scales of the order 
of the harmonic oscillator length $a_{HO}=(\hbar/m\omega)^{1/2}$.
Since the typical
size of the thermally excited cloud is of the order of the classical
thermal radius $(2k_BT/m\omega^2)^{1/2}$, the semiclassical approximation makes
sense only if $k_BT \gg \hbar \omega$ \cite{N2}.  
The sum over the quantum states $j$ labelling the solutions
of the Bogoliubov equations (\ref{bogeqs}) is replaced 
by the integral
$\int d{\bf p}/(2\pi\hbar)^3$.
By neglecting derivatives of $u$ and 
$v$ and second derivatives of $\varphi$ it is
possible to rewrite  
the Bogoliubov-type equations (\ref{bogeqs}) in the 
semiclassical form 
\begin{eqnarray}
\left( \frac{p^2}{2m} + V_{ext}({\bf r}) -\mu + 2gn({\bf r})
\right) u({\bf p},{\bf r}) + gn_0({\bf r})v({\bf p},{\bf r}) &=&
\epsilon({\bf p},{\bf r}) u({\bf p},{\bf r}) \;\; \nonumber \\
\label{bogeqs1} \\
\left( \frac{p^2}{2m} + V_{ext}({\bf r}) -\mu + 2gn({\bf r})
\right) v({\bf p},{\bf r}) + gn_0({\bf r})u({\bf p},{\bf r}) &=&
- \epsilon({\bf p},{\bf r}) v({\bf p},{\bf r}) \nonumber \;\;. 
\end{eqnarray}
The solution of equations (\ref{bogeqs1}) is easily obtained and one finds the
following result for the spectrum of the elementary excitations 
\begin{equation}
\epsilon({\bf p},{\bf r}) = \left( \left( \frac{p^2}{2m} + V_{ext}({\bf r})
- \mu + 2gn({\bf r}) \right)^2 - g^2n_0^2({\bf r}) \right)
^{1/2} \;\;,
\label{elex}
\end{equation}
with the coefficients
$u({\bf p},{\bf r})$ and $v({\bf p},{\bf r})$ given by
\begin{eqnarray}
&\;& u^2({\bf p},{\bf r}) = \frac{p^2/2m + V_{ext}({\bf r}) - \mu 
+ 2gn({\bf r})
+ \epsilon({\bf p},{\bf r})}{2\epsilon({\bf p},{\bf r})}
\nonumber \\
&\;& v^2({\bf p},{\bf r}) = \frac{p^2/2m + V_{ext}({\bf r}) - \mu 
+ 2gn({\bf r})
- \epsilon({\bf p},{\bf r})}{2\epsilon({\bf p},{\bf r})}
\label{bogeqs2} \\
&\;& u({\bf p},{\bf r})v({\bf p},{\bf r}) = - \frac{gn_0({\bf r})}
{2\epsilon({\bf p}, 
{\bf r})} \;\;. \nonumber 
\end{eqnarray}
The spectrum (\ref{elex}) for the elementary excitations is significant
only for energies $\epsilon({\bf p},{\bf r}) \gg \hbar \omega$. 
In fact the low-lying
modes of the system are not properly described in the semiclassical 
approximation and their study requires the solution of the 
coupled Bogoliubov equations (\ref{bogeqs}) \cite{OXF,HZG}.
As already mentioned above, the condition $\epsilon({\bf p},{\bf r})\gg\hbar
\omega$ restricts the study of thermodynamics
to temperatures much larger than the oscillator temperature $\hbar\omega/
k_B$. In practice the resulting analysis is useful only for systems
with large values of $N$, where the critical temperature is much larger than
the oscillator frequency (see eq.(\ref{Tc})).

It is interesting to analyze in more detail the excitation energies 
(\ref{elex}). First of all in the absence of interparticle interactions 
($g=0$) the elementary excitations reduce to the usual single particle 
spectrum in the presence of the external potential $V_{ext}$    
\begin{equation}
\epsilon^0({\bf p},{\bf r}) = \frac{p^2}{2m} + V_{ext}({\bf r}) - \mu \;\;,
\label{eps0}
\end{equation}
where the chemical potential is given by $\mu=3\hbar\bar{\omega}
/2$, and $\bar{\omega}=(\omega_x+\omega_y+\omega_z)/3$ is the 
arithmetic average of
the oscillator frequencies. In the limit $Na/a_{HO} \gg 1$ (Thomas-Fermi limit)
we can distinguish two
regions in space depending on whether the coordinate ${\bf r}$ lies 
inside or outside the condensate cloud. Inside the condensate region one 
can neglect 
the kinetic term in eq. (\ref{cequ}) and one has 
approximately $g(n_0+2n_T)=\mu-V_{ext}$. In this region  eq. 
(\ref{elex}) 
becomes   
\begin{equation}
\epsilon^{inside}({\bf p},{\bf r}) = \frac{p}{2m}\left( p^2 + 4mgn_0({\bf r}) 
\right)^{1/2} \;\;,
\label{epsin}
\end{equation}
which corresponds to the local-density form of the Bogoliubov spectrum.
For small $p$ eq. (\ref{epsin}) gives the phonon-like dispersion 
$\epsilon^{inside}({\bf p},{\bf r})=c({\bf r})p$ where $c({\bf r})=\sqrt{g
n_0({\bf r})/m}$ is the local velocity of sound.
Far from the condensate cloud one instead has $V_{ext}>>2g(n_0+n_T)$ and we are 
left with
\begin{equation}
\epsilon^{outside}({\bf p},{\bf r}) = \frac{p^2}{2m} + V_{ext}({\bf r}) 
- \mu \;\;,
\label{epsou}
\end{equation}
which has the same form as the independent particle spectrum (\ref{eps0}), 
{\it but} now the chemical potential contains the effects of interaction 
and is in general much larger than its non-interacting value.
Finally it is worth noting that when the energy 
$\epsilon({\bf p},{\bf r})$ is much larger than the chemical potential $\mu$
one obtains the Hartree-Fock type spectrum \cite{HF}
\begin{equation}
\epsilon^{HF}({\bf p},{\bf r}) = \frac{p^2}{2m} + V_{ext}({\bf r}) - \mu 
+2gn({\bf r}) \;\;.
\label{epshf}
\end{equation}
   
In the semiclassical scheme 
the canonical transformation (\ref{trans}) becomes the 
familiar
Bogoliubov transformation in momentum space relating the particle and 
quasi-particle creation and annihilation operators  
\begin{eqnarray}
\tilde{\psi}({\bf p},{\bf r}) &=& u({\bf p},{\bf r})\alpha({\bf p},{\bf r})
+ v({\bf p},{\bf r})\alpha^{\dagger}(-{\bf p},{\bf r}) \;\;,
\nonumber \\
\tilde{\psi}^{\dagger}({\bf p},{\bf r}) &=& u({\bf p},{\bf r})\alpha^{\dagger}
({\bf p},{\bf r}) + v({\bf p},{\bf r})\alpha(-{\bf p},{\bf r}) \;\;.
\label{trans1}
\end{eqnarray}
The quasi-particle distribution function $f({\bf p},
{\bf r})$ is obtained from the statistical average $\langle\alpha^{\dagger}
({\bf p},{\bf r})\alpha({\bf p},{\bf r})\rangle$, given by the usual Bose
distribution
\begin{equation}
f({\bf p},{\bf r}) = \langle\alpha^{\dagger}({\bf p},{\bf r})\alpha({\bf p},
{\bf r})\rangle = \frac{1}{\exp(\epsilon({\bf p},{\bf r})/k_BT) - 1} \;\;,
\label{qpdis}
\end{equation}
whereas the particle distribution function 
$F({\bf p},{\bf r})$ is given by the statistical average $\langle\tilde{\psi}
^{\dagger}({\bf p},{\bf r})\tilde{\psi}({\bf p},{\bf r})\rangle$ of the 
particle field operators in the coordinate-momentum representation. By making 
use of the transformation (\ref{trans1}) one gets
\begin{eqnarray}
F({\bf p},{\bf r}) = \langle\tilde{\psi}^{\dagger}({\bf p},{\bf r})\tilde{\psi}
({\bf p},{\bf r})\rangle &=& \left( u^2({\bf p},{\bf r}) + v^2({\bf p},{\bf r}) 
\right) f({\bf p},{\bf r}) \nonumber \\ 
&=& - \left(\frac{\partial\epsilon({\bf p},{\bf r})}
{\partial\mu}\right)_{n_0,n_T} f({\bf p},{\bf r}) \;\;,
\label{pdis}
\end{eqnarray}
with
\begin{equation}
\left(\frac{\partial\epsilon({\bf p},{\bf r})}
{\partial\mu}\right)_{n_0,n_T} = - \; \frac{p^2/2m+V_{ext}({\bf r})-\mu+
2gn({\bf r})}{\epsilon({\bf p},{\bf r})} \;\;.
\label{Ftof}
\end{equation}
It is worth noticing that
the particle and quasi-particle distribution functions differ only in the 
low-energy regime. In fact at high energies the excitation spectrum 
takes the HF form (\ref{epshf}), $\partial\epsilon/\partial\mu = -1$ and 
hence $F({\bf p},{\bf r})=f({\bf p},{\bf r})$. Conversely in the phonon regime 
one has $F({\bf p},{\bf r})=mc({\bf r}) f({\bf p},{\bf r})/p$.
    
The non-condensate density $n_T$ can be obtained in a self consistent way
by integrating over momenta the particle distribution (\ref{pdis}) 
\begin{equation}
n_T({\bf r}) = \int\;\frac{d{\bf p}}{(2\pi\hbar)^3} F({\bf p},{\bf r}) \;\;,
\label{nT1}
\end{equation}
and finally the chemical potential $\mu$ is fixed by the normalization 
condition
\begin{equation}
N = N_0(T) + N_T = \int\;d{\bf r} n_0({\bf r}) + \int\;d{\bf r} n_T({\bf r})  
\;\;,
\label{norm}
\end{equation}
where $N_0(T)$ and $N_T$ are respectively the total numbers of atoms {\it in} 
the condensate and {\it out} of the condensate in equilibrium 
at temperature $T$.

Equations  (\ref{cequ}), (\ref{elex}) and (\ref{qpdis})-(\ref{norm}) must be 
solved self-consistently. The solution of these equations yields directly,
for a given temperature $T$, the chemical potential $\mu$,  
the condensate density 
$n_0({\bf r})$ and 
the density of thermally excited atoms $n_T({\bf r})$. 

Starting from the results discussed above it is possible to study the 
thermodynamic properties of the system whose excited states can be classified 
in terms of the states of a gas of undamped quasi-particles  
with energy spectrum (\ref{elex}).
The entropy of the system is then readily obtained through the combinatorial  
formula \cite{LL}
\begin{equation}
\frac{S}{k_B} =    
\int\;\frac{d{\bf r}d{\bf p}}{(2\pi\hbar)^3} \left( 
\frac{\epsilon({\bf p},{\bf r})}{k_BT} 
f({\bf p},{\bf r}) - \log(1-\exp(-\epsilon({\bf p},{\bf r})/k_BT)) \right) \;\;.
\label{stot}
\end{equation}
One must notice that the dependence of the entropy on $T$ is determined also
through the temperature dependences of the quantities $\mu$, $n_0$ and $n_T$ 
which enter the excitation spectrum $\epsilon({\bf p},{\bf r})$ (see 
eq. (\ref{elex})).
Starting from the entropy one can work out all the thermodynamic quantities. 
For example the total energy of a $N$-particle system at temperature $T$ 
can be calculated using the
thermodynamic relation 
\begin{equation}
E(T) = E(T=0) + \int_0^T\;dT'\; C_N(T')\;\;,
\label{et}
\end{equation}
where 
\begin{equation}
C_N(T)=T\left( \frac{\partial S}{\partial T} \right)_N 
\label{cN}
\end{equation}
is the specific heat at constant $N$ and $E(T=0)$ is the energy at zero 
temperature given by the well known Gross-Pitaevskii energy 
functional \cite{OX1,DS}.
\begin{equation}
E(T=0)=\int\;d{\bf r} \Phi({\bf r}) \left( -\frac{\nabla^2}{2m} + 
V_{ext}({\bf r}) + \frac{g}{2}n_0({\bf r}) \right)\Phi({\bf r}) \;\;,
\label{GPen}
\end{equation}
where $\Phi({\bf r})$ is here the condensate wave function at $T=0$.   

In the last part of this section we discuss the 
rotational properties of a trapped Bose gas which point out the 
effects of superfluidity \cite{STR}. 
The moment of inertia $\Theta$, 
relative to the 
$z$ direction, is defined as the linear response of the system to 
the perturbation field $-\Omega L_z$ according to the relation
\begin{equation}
\langle L_z \rangle = \Omega\Theta \;\;,
\label{momin}
\end{equation}
where $L_z$ is the $z$-component of the angular momentum. In terms of 
the particle field operator one has
\begin{equation}
L_z = \int \;d{\bf r} \psi^{\dagger}({\bf r}) 
\; {\ell}_z \; \psi({\bf r}) \;\;,
\label{Lz}
\end{equation}
where ${\ell}_z=-i\hbar (x\partial/\partial y - y\partial/\partial x)$.
The statistical average in eq. (\ref{momin}) must be calculated in the 
ensemble relative to 
the perturbed grand-canonical Hamiltonian $K'(\Omega) = 
K-\Omega L_z$.  
In the frame rotating with angular velocity
$\Omega$ 
the quasi-particles have energies 
\begin{equation}
\epsilon_{\Omega}({\bf p},{\bf r}) = \left( \left( \frac{p^2}{2m} + 
V_{ext}({\bf r})
- \mu + 2gn({\bf r}) \right)^2 - g^2n_0^2({\bf r}) \right)
^{1/2} - \Omega ({\bf r}\times {\bf p})_z \;\;,
\label{epsom}
\end{equation}
and are distributed according to the Bose function 
\begin{equation}
f_{\Omega}({\bf p},{\bf r}) = \frac{1}{\exp(\epsilon_{\Omega}
({\bf p},{\bf r}))-1} \;\;.
\label{fom}
\end{equation}
In the limit of an axially symmetric trap ($\omega_x\rightarrow\omega_y$), the 
condensate wave function does not contribute to the average 
$\langle L_z\rangle$ which is determined only by the angular momentum carried 
by the elementary excitations: 
\begin{equation}
\langle L_z\rangle = \int\;\frac{d{\bf r}d{\bf p}}{(2\pi\hbar)^3}
({\bf r}\times{\bf p})_z f_{\Omega}({\bf p},{\bf r}) \;\;.
\label{amom}
\end{equation}   
To lowest order in $\Omega$ one obtains the following relevant 
result for the moment of inertia $\Theta$
\begin{equation}
\Theta = - \int\; \frac{d{\bf r}d{\bf p}}{(2\pi\hbar)^3} 
\; ({\bf r}\times{\bf p})_z^2 \; \frac{\partial f({\bf p},{\bf r})}
{\partial\epsilon({\bf p},{\bf r})}
\;\;.
\label{momin1}
\end{equation}
Eq. (\ref{momin1}) can be also written in the useful form 
\begin{equation}
\Theta = \int\;d{\bf r} (x^2+y^2) \rho_n({\bf r}) \;\;,
\label{momin2}
\end{equation}
where 
\begin{equation}
\rho_n({\bf r}) = - \int\;\frac{d{\bf p}}{(2\pi\hbar)^3}\; \frac{p^2}{3} \;
\frac{\partial f({\bf p},{\bf r})}{\partial\epsilon({\bf p},{\bf r})} 
\label{rhon}
\end{equation}
provides a generalization of the most famous Landau formula for the density 
of the normal component to non homogeneous systems.
The deviation of the moment of inertia from the rigid value 
\begin{equation}
\Theta_{rig} = m \int\;d{\bf r} (x^2 + y^2) n({\bf r}) \;\;,
\label{momrig}
\end{equation}
provides a "measure" 
of the superfluid behaviour of the system (see section 3-C).

\subsection{Hartree-Fock approximation}

In the previous section we have seen that outside the condensate and in 
general for high excitation energies 
the Popov quasi-particle spectrum (\ref{elex}) reduces to the  
familiar Hartree-Fock 
spectrum (\ref{epshf}). 
The self-consistent HF method has been already employed to study the 
thermodynamic properties of
inhomogeneous dilute Bose systems \cite{HF,SNS}.  

The HF approximation can be easily obtained by neglecting the small ''hole" 
components $v_j$ in the first of the Bogoliubov equations (\ref{bogeqs}). The 
resulting equation then takes the form
\begin{equation}
{\cal L} u_j({\bf r}) \equiv \left(-\frac{\nabla^2}{2m} + V_{ext}({\bf r})
- \mu +2gn({\bf r}) \right) u_j({\bf r}) = \epsilon_j u_j({\bf r}) \;\;.
\label{hfeq}
\end{equation}
Notice that the Hamiltonian (\ref{hfeq}) differs from the effective Hamiltonian (\ref{cequ}) yielding the condensate wave function (for a more detailed 
derivation and discussion of the Hartree-Fock approximation for  
Bose gases see Ref. \cite{HF}).
By applying the semiclassical prodedure to solve
eq. (\ref{hfeq}), one finds the energy spectrum $\epsilon_
{HF}({\bf p},{\bf r})$ of eq. (\ref{epshf}).  
In the HF approximation there is no difference between particles 
and quasi-particles and therefore the
distribution functions $f({\bf p},{\bf r})$ and 
$F({\bf p},{\bf r})$ coincide ($\partial\epsilon_{HF}/\partial\mu = -1$)
\begin{equation}
F({\bf p},{\bf r}) = f({\bf p},{\bf r}) = \frac{1}{\exp(\epsilon_{HF}({\bf p},
{\bf r})/k_BT) - 1} \;\;.
\label{pdhf}
\end{equation}

Results (\ref{stot})-(\ref{cN}) for the thermodynamic quantities obtained in 
section 2-A hold also in 
the HF approximation if one simply replaces the excitation spectrum $\epsilon(
{\bf p},{\bf r})$ with the corresponding HF spectrum (\ref{epshf}).
By calculating the statistical average of the grand-canonical Hamiltonian 
(\ref{gch}) over the eigenstates eq. (\ref{hfeq})
one obtains the following explicit expression for the energy of the system 
in the HF approximation
\begin{eqnarray}
E_{HF}&=&\int\;d{\bf r} \Phi({\bf r})\left(-\frac{\nabla^2}{2m}\right)
\Phi({\bf r}) + \int\;\frac{d{\bf r}d{\bf p}}{(2\pi\hbar)^3} \frac{p^2}{2m}
F({\bf p},{\bf r}) \nonumber \\ 
&+& \int\;d{\bf r} V_{ext}({\bf r}) n({\bf r}) + \frac{g}{2}
\int\;d{\bf r} (2n^2({\bf r}) - n_0^2({\bf r})) \;\;.
\label{hfen}
\end{eqnarray}
In the above equation the first two terms give respectively the kinetic 
contribution from the condensate and non-condensate particles. 
The third term
gives the confining energy averaged over the total density $n$, and the last
term represents the interaction contribution.  
The HF energy functional (\ref{hfen}) 
is consistent with the thermodynamic
relations ((\ref{stot})-(\ref{cN})).  
 
Results (\ref{momin1})-(\ref{rhon}) for the moment of inertia 
also hold in the HF 
approximation with the distribution 
$f({\bf p},{\bf r})$ given by (\ref{pdhf}).  
In the HF approximation one easily finds $\rho_n({\bf r})=m n_T({\bf r})$ and 
hence the moment of inertia is given by 
\begin{equation}
\Theta_{HF} = m \int\;d{\bf r}\;(x^2+y^2)\;n_T({\bf r}) 
\label{momhf}
\end{equation}
consistently with the finding of Ref. \cite{STR}. The identification of the 
normal component $\rho_n$ with the non-condensate density is a peculiar 
property of the HF approximation.  
In the Popov approach these two quantities 
behave differently at low temperatures because of the presence of phonon-type 
excitations.  

\subsection{The non-interacting model}

The results for the thermodynamic functions of the trapped ideal gas are 
easily obtained by setting the coupling constant $g$ equal to zero in the
equations derived above. The equation for the condensate 
(\ref{cequ}) takes the form
\begin{equation}
\left( -\frac{\nabla^2}{2m} + V_{ext}({\bf r}) - \mu \right)\Phi({\bf r}) = 0
\;\;,
\label{cequ0}
\end{equation}
whose solution is the gaussian function  
\begin{equation}
\Phi({\bf r}) = \sqrt{N_0}\left(\frac{m\omega}{\pi\hbar}\right)^{3/4}
\exp\left(-\frac{m}{2\hbar}(\omega_x x^2+\omega_y y^2 + \omega_z z^2)\right) 
\;\;,
\label{gauss}
\end{equation}
normalized to the number of condensate atoms $\int d{\bf r} |\Phi|^2 = N_0$, 
whereas the chemical potential is $\mu=\hbar(\omega_x+\omega_y+\omega_z)/2$.
The semiclassical spectrum (\ref{elex}) takes the form (\ref{eps0}) and the 
non-condensate density is given by 
\begin{eqnarray}
n_T({\bf r})&=&\int\;\frac{d{\bf p}}{(2\pi\hbar)^3}(\exp((p^2/2m
+V_{ext}({\bf r})
-\mu)/k_B T) -1)^{-1} 
\nonumber\\
&=& \lambda_T^{-3} g_{3/2}(\exp((V_{ext}({\bf r})-\mu)/k_BT))
\;\;,
\label{nT0}
\end{eqnarray}
where $\lambda_T=\hbar\sqrt{2\pi/mk_BT}$ is the thermal wavelength and 
$g_\alpha(z)=\sum_{n=1}^{\infty}z^n/n^\alpha$ is the usual Bose function.
The total number of non-condensate atoms is obtained by integrating $n_T$ over 
space
\begin{equation}
N_T = \int\;d{\bf r}\;n_T({\bf r}) =\left(\frac{k_BT}{\hbar\omega}\right)^3
g_3(\exp(\mu/k_BT)) \;\;.
\label{NT0}
\end{equation}
In the region of temperatures where the semiclassical approximation for the
excited states is valid, $k_BT\gg \hbar\omega$, one can expand 
the $g_3(z)$ function 
around $z=1$ and obtain to lowest order in $N^{-1/3}$
\begin{equation}
\frac{N_T}{N} = t^3 + \frac{3\zeta(2)}
{2(\zeta(3))^{2/3}} \frac{\bar{\omega}}{\omega} N^{-1/3} t^2
\;\;,
\label{NT01}
\end{equation}
where we have introduced the reduced temperature $t=T/T_c^0$, 
and $T_c^0=(N/\zeta(3))^{1/3}\hbar\omega/k_B$ is the critical temperature 
in the 
large $N$ limit. Finite size effects, which are accounted for by the second 
term in eq. (\ref{NT01}), are proportional to the ratio 
$\bar{\omega}/\omega$ between
the arithmetic $\bar{\omega}=(\omega_x+\omega_y+\omega_z)/3$ and the geometric
averages of the oscillator frequencies and thus depend on the trap anisotropy
parameter $\lambda=\omega_z/\omega_x=\omega_z/\omega_y$, having a   
minimum for an isotropic trap ($\lambda=1$). These finite      
size effects shift the value of the critical temperature from $T_c^0$ 
towards lower temperatures. By setting the right hand side of eq. (\ref{NT01})
equal to unity one finds the following result for the shift in the critical 
temperature \cite{IGAS}
\begin{equation}
\frac{\delta T_c^0}{T_c^0}=- \frac{\zeta(2)}{2(\zeta(3))^{2/3}}
\frac{\bar{\omega}}
{\omega}N^{-1/3}\simeq - 0.73 \frac{\bar{\omega}}{\omega}N^{-1/3} \;\;.
\label{dtc}
\end{equation}
The temperature dependence of the energy per particle  
is easily obtained in the semiclassical approximation and one has
(for $t<1$) 
\begin{equation}
\frac{E}{Nk_BT_c^0} = \frac{3\zeta(4)}{\zeta(3)}t^4 + 3(\zeta(3))^{1/3}
\frac{\bar{\omega}}{\omega} N^{-1/3} t^3 + \frac{3(\zeta(3))^{1/3}}{2}
\frac{\bar{\omega}}{\omega} N^{-1/3} \;\;.
\label{E001}
\end{equation}
It is worth noticing that the shift (\ref{dtc}) in the critical temperature 
as well as the  $N^{-1/3}$ terms in eqs. (\ref{NT01}), 
(\ref{E001}) exactly coincide with the results obtained to lowest order in 
$N^{-1/3}$ from a full quantum treatment of the non-interacting model in the 
large $N$ limit \cite{IGAS}. For $N=1000$ the predictions 
(\ref{NT01}), (\ref{E001}) 
are already indistinguishable from the exact result obtained 
by summing explicitly over the excited states of the 
harmonic oscillator Hamiltonian. This is well illustrated in Fig. 1 where we 
plot the prediction $N_0/N$ (see eq. (\ref{NT01})) for the temperature 
dependence of the 
condensate fraction compared to the exact quantum calculation and the large
$N$ behaviour  
$N_0/N=1-t^3$ obtained by setting $\mu=0$ in eq. (\ref{NT0}).
The accuracy of the semiclassical approximation reveals that the main origin 
of finite size effects in the thermodynamic properties of these systems 
arises from the quantum effects in the equation for the 
condensate, which are responsible for the proper value of the chemical 
potential, rather than from the quantum effects in the equation for the 
excited states. 

It is also important to notice that in the large $N$ limit all the 
thermodynamic properties in the non-interacting model depend  
only on the reduced temperature $t=T/T_c^0$
(see eqs. (\ref{NT01}), (\ref{E001}) with $N\rightarrow\infty$). In the next 
section it will be shown that in the presence of interaction  
the thermodynamic properties of the system depend also on  
the additional parameter 
$\eta$ defined in eq. (\ref{ratio1}).
 
Finally, the ratio $\Theta/\Theta_{rig}$ of the moment of inertia to its 
rigid value has been calculated for the non-interacting model in Ref. 
\cite{STR}. In terms of the reduced temperature $t$ this result reads
\begin{equation}
\frac{\Theta}{\Theta_{rig}} = 1 - \frac{1-t^3}{1 - t^3 + \gamma t^4 (\lambda N)
^{1/3}} 
\;\;,
\label{mi0}
\end{equation}
where $\gamma=2\zeta(4)/(\zeta(3))^{4/3}\simeq1.69$. Starting from  (\ref{mi0})
one finds that 
in the non-interacting model the effects of finite size are much more 
important for $\Theta$ than for the other thermodynamic quantities.

\subsection{Thomas-Fermi limit and scaling}

When the number of atoms in the condensate is enough large the Thomas-Fermi 
approximation, which is obtained neglecting the quantum kinetic energy term 
in the equation for the condensate, turns out to be very accurate 
\cite{OX1,BP,DS}. In this limit the Popov 
mean-field equations show an important scaling behaviour that we have recently
investigated in Ref. \cite{US1}. The scaling behaviour is best illustrated by
introducing the dimensionless quantities
$\tilde{p}_i = \sqrt{1/2mk_BT_c^0}p_i$, $\tilde{r}_i=\sqrt{m/2k_BT_c^0}
\omega_i r_i$,
$\tilde{\mu} = \mu/k_BT_c^0$, $\tilde{\epsilon} = \epsilon/k_BT_c^0$, 
$\tilde{n}_T=n_T(2k_BT_c^0/m\omega^2)^{3/2}/N$,  
$\tilde{n}_0=n_0(2k_BT_c^0/m\omega^2)^{3/2}/N$ and the reduced temperature 
$t=T/T_c^0$.  
By neglecting the quantum kinetic energy term in eq. (\ref{cequ}) the  
mean-field equations can be rewritten in the following reduced form
\begin{equation}
\tilde{n}_0(\tilde{r}) = \frac{1}{\tilde{g}} \left( \tilde{\mu} 
- \tilde{r}^2 - 
2\tilde{g}\tilde{n}_T \right) 
\;\theta(\tilde{\mu}-\tilde{r}^2-2\tilde{g}\tilde{n}_T) \;\;,
\label{red1}
\end{equation}
\begin{equation}
\tilde{\epsilon}(\tilde{p},\tilde{r}) = \sqrt{\left(\tilde{p}^2
+\tilde{r}^2-\tilde{\mu}
+2\tilde{g}(\tilde{n}_0+\tilde{n}_T)\right)^2
-\tilde{g}^2\tilde{n}_0^2} \;\;, 
\label{red2}
\end{equation}
and
\begin{equation}
\tilde{n}_T(\tilde{r}) =  
\frac{1}{\pi^3\zeta(3)}\int\; d\tilde{{\bf p}} 
\left(-\frac{\partial\tilde{\epsilon}}{\partial\tilde{\mu}}\right)
f(\tilde{\epsilon}/t) \;\;,
\label{red3}
\end{equation}
where the renormalized coupling constant $\tilde{g}$ is given by 
\begin{equation}
\tilde{g}=8\pi\eta^{5/2}/15 \simeq 5.17 \left(N^{1/6}\frac{a}{a_{HO}}
\right) \;\;,
\label{gtilde}
\end{equation}
and is hence fixed by the interaction parameter $\eta$ defined in 
(\ref{ratio1}).
In eq. (\ref{gtilde})     
$\theta(x)$ is the step function and $f(\tilde{\epsilon}/t)=(\exp(\tilde
{\epsilon}/t)-1)^{-1}$ is the Bose function in terms of the reduced variables. 
The normalization condition for the reduced densities reads:
$\int d\tilde{\bf r} (\tilde{n}_0+\tilde{n}_T) = 1$.
Equations (\ref{red1})-(\ref{red3}) exhibit the anticipated 
scaling behaviour in the variables $t$ and $\eta$ \cite{N3}. Starting from 
their 
solutions
one can calculate 
the condensate and the thermal densities as well as the 
excitation spectrum. This gives access to 
all the relevant thermodynamic quantities of the system.
For example the condensate fraction is given by
\begin{equation}
{N_0 \over N} = 1 - 
\frac{1}{\pi^3\zeta(3)}\int\; d\tilde{{\bf r}} d\tilde{{\bf p}} 
\left(-\frac{\partial\tilde{\epsilon}}{\partial\tilde{\mu}}\right)
f(\tilde{\epsilon}/t) \;\;.
\label{N0red}
\end{equation}
In particular the vanishing of the right hand side of eq. (\ref{N0red}) fixes 
the value of the
critical temperature $t_c=T_c/T_c^0$.
To the lowest order in the coupling constant $\tilde{g}$ one finds \cite{GPS}
\begin{equation}
t_c \simeq 1 - 0.43 \eta^{5/2} = 1-1.3 N^{1/6}a/a_{HO} \;\;.
\label{dtc1}
\end{equation}
The energy of the system can be calculated starting
from the standard thermodynamic relation in terms of the entropy (see eq. 
(\ref{et})).
In units of $Nk_BT_c^0$ one finds:
\begin{equation}
{E\over Nk_BT_c^0} = \frac{5}{7}\eta + \int_0^t\;dt' t'\frac{\partial s}
{\partial t'} \;\;, 
\label{Ered}
\end{equation}
where the entropy per particle $s$ is given by (see eq. (\ref{stot})) 
\begin{eqnarray}
s(t,\eta) &=& k_B \frac{1}{\pi^3\zeta(3)} \int\;
d\tilde{{\bf r}}d\tilde{{\bf p}} \left( \frac{\tilde{\epsilon}}{t}
f(\tilde{\epsilon}/t) - \log(1-\exp(-\tilde{\epsilon}/t)) \right)
\;\;.
\label{sred}
\end{eqnarray}

Finally, the ratio $\Theta/\Theta_{rig}$ of the moment of inertia to its 
rigid value (see eqs. (\ref{momin1}), (\ref{momrig})) takes the form 
\begin{equation}
\frac{\Theta}{\Theta_{rig}}=\frac{2}{3\pi^3\zeta(3)}\frac{1}{t} \frac
{\int\;d\tilde{{\bf r}}d\tilde{{\bf p}}(\tilde{x}^2+\tilde{y}^2)\tilde{p}^2
f(\tilde{\epsilon}/t) (1+f(\tilde{\epsilon}/t)}
{\int\;d\tilde{{\bf r}} (\tilde{x}^2+\tilde{y}^2) (\tilde{n}_0(\tilde{{\bf r}})
+\tilde{n}_T(\tilde{{\bf r}}))} \;\;.
\label{totrig}
\end{equation}

In the next sections we will show that the scaling behaviour is well achieved 
for the choice of parameters corresponding to the available experimental 
conditions. Scaling is consequently expected to provide a powerful tool for a 
systematic investigation of these trapped Bose gases.

\subsection{Elementary excitations and low temperature behaviour}

In this section we discuss the thermodynamic behaviour of a trapped Bose gas 
at low temperature ($k_BT\ll\mu$). Though at present this regime is not 
easily available in experiments, its theoretical investigation is highly 
interesting and reveals unexpected features concerning the competition 
between collective and single-particle effects which are worth discussing.

We will derive our results in the framework of the scaling regime discussed in
section 2-D, where one assumes that the number of atoms in the trap is large
enough to justify the use of the Thomas-Fermi approximation for the condensate 
and that the excited states can be described in the semiclassical 
approximation. This allows one to write the relevant thermodynamic functions 
in the form of integrals in coordinate and momentum space 
(see eqs. (\ref{N0red}), (\ref{Ered})-(\ref{totrig}))
which are well suited for an analytic expansion at low temperature. This 
expansion corresponds to exploring the region where $k_BT\ll\mu$. At the same 
time the thermal energy $k_BT$ must be larger than the typical quanta of 
energy characterizing the elementary excitations of the system, in order to 
justify the use of the semiclassical approximation. In the absence of two-body
interactions the quantum of energy is fixed by the oscillator energy $\hbar
\omega$. In the presence of interactions one has to distinguish between two
regions. Inside the condensate the elementary excitations are of collective
nature. For a spherical trap the dispersion law of these excitations can be 
expressed analytically in terms of the angular momentum quantum number $\ell$ 
and of the 
number of radial nodes according to the law \cite{STR1} 
\begin{equation}
\epsilon(n,\ell) = \hbar\omega (2n^2+2n\ell+3n+\ell)^{1/2} \;\;.
\label{Stex}
\end{equation}
Their typical energy separation is still of the order of the oscillator 
energy $\hbar\omega$. Near the classical boundary the relevant excitations 
are instead of single-particle type and in first approximation can be 
described by the 
Hartree-Fock Hamiltonian (\ref{hfeq}) which in the Thomas-Fermi limit, and 
for spherical traps takes the 
simplified form
\begin{equation}
{\cal L}_{TF} = -\frac{\nabla^2}{2m} + \frac{m}{2}\omega^2 |r^2-R^2| \;\;,
\label{HHF}
\end{equation}
where $R=\sqrt{2\mu/m\omega^2}$ is the classical boundary of the condensate. 
Actually the Hartree-Fock Hamiltonian (\ref{HHF}) gives correctly the 
energy 
of the elementary excitations only for $r\ge R$ (see eq. (\ref{epsou})). 
In fact for $r<R$
it ignores Bogoliubov effects. 
The potential term of the Hamiltonian (\ref{HHF}) tends to confine the low 
energy 
single-particle states near the boundary. The typical energy gaps
characterizing the low energy spectrum of (\ref{HHF}) are of the order 
$\hbar\omega(R/a_{HO})^{2/3}\sim \hbar\omega(\mu/\hbar\omega)^{1/3}$. 
In the large $N$ limit this energy is larger than the oscillator energy, but 
still smaller than $\mu$. As a consequence one can find a useful range of 
temperatures where to apply the low $t$ expansion. 
In a homogeneous gas this region would correspond to the regime of collective 
excitations (phonons) which entirely determine the thermodynamics of the 
system. In a trapped Bose gas both collective and single-particle excitations 
taking place near the classical boundary
can be important in the determination of the low temperature behaviour. 
Actually, as it will be shown, for most thermodynamic quantities the low 
$t$ behaviour is governed by single-particle excitations. 

Let us first consider the thermal energy of the system. 
At low temperature one can 
neglect the $t$-dependence in the interaction term characterizing the 
excitation spectrum (\ref{red2}) and the integral appearing in (\ref{Ered}) 
can be calculated
explicitly giving the result 
\begin{eqnarray}
\frac{E}{Nk_BT_c^0} = \frac{5}{7}\eta 
&+& \frac{1}{\pi^3\zeta(3)}\int_{\tilde{r}<\sqrt{\tilde{\mu}_0}}\;
d^3\tilde{{\bf p}}
d^3\tilde{{\bf r}} \frac{\sqrt{\tilde{p}^4+2\tilde{g}
\tilde{n}_0\tilde{p}^2}}{\exp(\sqrt{\tilde{p}^4+2\tilde{g}\tilde{n}_0
\tilde{p}^2}/t)-1} \nonumber\\
&+& \frac{1}{\pi^3\zeta(3)}\int_{\tilde{r}\ge\sqrt{\tilde{\mu}_0}}\;
d^3\tilde{{\bf p}}
d^3\tilde{{\bf r}} \frac{\tilde{p}^2+\tilde{r}^2-\tilde{\mu}_0}
{\exp((\tilde{p}^2+\tilde{r}^2-\tilde{\mu}_0)/t)-1} \;\;,
\label{eterm}
\end{eqnarray}
where we have used the dimensionless variables of section 2-D with 
$\tilde{\mu}_0=\mu(T=0)/k_BT_c^0=\eta$ which at $T=0$ fixes the size of the 
condensate $\tilde{R}=\sqrt{\tilde{\mu}_0}$ (see eq. (\ref{red1})). 
The first integral on the right hand side (r.h.s.) of eq. (\ref{eterm})
gives the contribution to the energy from the region inside the condensate 
where the Bogoliubov effects take place, the second integral yields 
the contribution from the region outside the condensate where the spectrum 
is particle-like.  
By taking the 
linear phonon approximation $\tilde{\epsilon}=\tilde{p}\sqrt{2\tilde{g}
\tilde{n}_0}$ for the excitation spectrum one finds that the first integral 
on the r.h.s. of (\ref{eterm}) has a
divergent behaviour near the boundary $\tilde{R}$ as a consequence of the 
vanishing 
of the local sound velocity. This rules out the typical $t^4$ phonon-type 
behaviour for the temperature dependence of the energy. In fact a careful 
expansion of the integral (\ref{eterm}) at low $t$ shows that the relevant 
region contributing to the integral is near the boundary. 
The expansion yields the law
\begin{equation}
\frac{E}{Nk_BT_c^0}=\frac{5}{7}\eta + A \sqrt{\eta} t^{7/2} \;\;,
\label{eexp}
\end{equation}
where $A$ is a numerical factor.
This behaviour points out the 
single-particle nature of the leading contribution to the thermal energy.
In terms of the density of states  
one finds the result, valid for 
$\epsilon\rightarrow 0$, $g(\epsilon)\sim\epsilon^{3/2}$
which should be compared with the typical phonon law $g(\epsilon)\sim
\epsilon^2$.

An opposite result is obtained if one investigates the low temperature 
behaviour of the thermal depletion (see eq. (\ref{N0red}))
\begin{eqnarray}
\frac{N_T}{N} &=&
\frac{1}{\pi^3\zeta(3)}\int_{\tilde{r}<\sqrt{\tilde{\mu}_0}}\;
d^3\tilde{{\bf p}}
d^3\tilde{{\bf r}} \frac{\tilde{p}^2+\tilde{g}\tilde{n}_0}
{\sqrt{\tilde{p}^4+2\tilde{g}
\tilde{n}_0\tilde{p}^2}}\frac{1}{\exp(\sqrt{\tilde{p}^4+2\tilde{g}\tilde{n}_0
\tilde{p}^2}/t)-1} \nonumber \\ 
&+& \frac{1}{\pi^3\zeta(3)}\int_{\tilde{r}\ge\sqrt{\tilde{\mu}_0}}\;
d^3\tilde{{\bf p}}
d^3\tilde{{\bf r}} \frac{1}
{\exp((\tilde{p}^2+\tilde{r}^2-\tilde{\mu}_0)/t)-1} \;\;.
\label{dterm}
\end{eqnarray}
In fact in this case the particle distribution function $F({\bf p},{\bf r})$, 
whose integral gives the thermal depletion, is enhanced at small $p$ by the 
phonon contribution (see eq. (\ref{Ftof})) which provides the leading effect in 
the integral (\ref{dterm}) according to the law: 
\begin{equation}
\frac{N_T}{N}=\frac{\pi^2}{3\sqrt{2}\zeta(3)} \eta t^2 \;\;.
\label{dexp}
\end{equation}
Eq. (\ref{dexp}) provides the generalization of the Ferrel law \cite{FER} 
holding for Bose superfluids to a trapped 
Bose gas. In this case the single particle contribution arising from the 
boundary region is a higher order effect and behaves as $t^{5/2}$.

In the scaling limit one can also obtain a useful estimate for the quantum 
depletion of the condensate. At $T=0$ the density of non-condensate atoms 
is given, in the 
semiclassical approximation, by the integral  
$\delta n({\bf r}) = \int d{\bf p}
/(2\pi\hbar)^3 v^2({\bf p},{\bf r})$, where $v^2({\bf p},{\bf r})$ is given 
in eq. (\ref{bogeqs2}). By employing the Thomas-Fermi approximation  
one finds the result
\begin{equation}
\delta\tilde{n}(\tilde{r}) = \frac{2\sqrt{2}}{3\pi^2\zeta(3)}
(\tilde{\mu}_0-\tilde{r}^2)^{3/2} \theta(\tilde{\mu}_0-\tilde{r}^2) \;\;,
\label{deln}
\end{equation}
already discussed in Ref. \cite{HUA}.
It is worth noticing that the relative local quantum depletion is 
given by
$\delta n({\bf r})/n({\bf r})=8/3(a^3n({\bf r})/\pi)^{1/2}$ which coincides 
with the well known result for the quantum depletion of a homogeneous Bose 
gas \cite{BOG}. 
By integrating (\ref{deln}) over space one gets the total number of 
quantum depleted 
atoms
\begin{equation}
\frac{\delta N}{N} = \int\;d\tilde{{\bf r}}\delta\tilde{n}(\tilde{r})
= \frac{\eta^3}{6\sqrt{2}\zeta(3)} \simeq 0.1 \eta^3 \;\;.
\label{qdepl}
\end{equation}
The above estimate of the quantum depletion agrees reasonably well with 
the numerical results 
of Ref. \cite{HZG} obtained from the full solution of the Bogoliubov equations.

Let us finally discuss the low temperature behaviour of the 
moment of inertia. The normal component (\ref{totrig}) 
gets a contribution from 
both the collective and single-particle excitations. One can easily show that
the leading contribution to the integral (\ref{totrig}) 
at low temperature is given 
by single-particle excitations and behaves as
\begin{equation}
\frac{\Theta}{\Theta_{rig}} \sim  t^{5/2} \;\;.
\label{mexp}
\end{equation}
In this case the phonon contribution is a higher order effect and 
behaves as $t^4$.

\section{Results}

In this section we present the numerical results obtained by solving the  
equations of the Popov approximation both below and above 
the critical temperature. 
In section 3-A we discuss 
the temperature dependence of the condensate fraction 
and of the peak density in the center of the trap. Other structural 
properties such as the temperature dependence of the density profiles and of 
the mean square radii of the
condensed and non-condensed part of the cloud are discussed. 
In section 3-B we present the results for the temperature 
dependence of the energy and of the specif heat 
Finally in section 3-C 
we discuss the temperature dependence of the moment of inertia.      

In the condensed phase, $T<T_c$, the self-consistent procedure is 
schematically divided in the following steps: 

i) the equation for the condensate wave-function
\begin{equation}
\left[ - \frac{\nabla^2}{2m} + V_{ext}({\bf r}) - \mu 
+ g(n_0({\bf r}) + 2n_T({\bf r}))\right] \Phi({\bf r}) = 0 
\label{cequ1}
\end{equation}
is solved, using the method described in Ref. \cite{DS}, for the condensate 
density $n_0({\bf r})$ and the chemical potential $\mu$, while keeping fixed 
the number of particles in the condensate, $N_0(T)$, and the density $n_T
({\bf r})$ of thermally excited atoms.

ii) The condensate density and the chemical potential found above are used to
calculate the excitation energies
\begin{equation}
\epsilon({\bf p},{\bf r}) = \left( \left( \frac{p^2}{2m} + V_{ext}({\bf r})
- \mu + 2gn({\bf r}) \right)^2 - g^2n_0^2({\bf r}) \right)
^{1/2} \;\;,
\label{elex1}
\end{equation}
which yield a new non-condensate density $n_T$ through the equation
(see eqs. (\ref{qpdis}), (\ref{pdis}))    
\begin{equation}
n_T({\bf r}) = \int\;\frac{d{\bf p}}{(2\pi\hbar)^3}
\left(-\frac{\partial\epsilon({\bf p},{\bf r})}
{\partial\mu}\right)_{n_0,n_T} (\exp(\epsilon({\bf p},{\bf r})/k_BT) 
- 1))^{-1} \;\;.
\label{dT1}
\end{equation}

iii) From the thermal density $n_T$ the total number of non-condensed atoms
is obtained by direct integration
\begin{equation}
N_T = \int\;d{\bf r} n_T({\bf r}) \;\;,
\label{NT1}
\end{equation}
and a new value for the number of atoms in the condensate is obtained from the 
normalization condition
\begin{equation}
N=N_0(T) + N_T \;\;.
\label{norm1}
\end{equation}

iv) With the new density $n_T$ and the new value for $N_0(T)$ steps (i)-(iii) 
are repeated until convergence is reached. We stop the iterative procedure
when successive iterations 
give values for the chemical potential and the condensate fraction, $N_0(T)/N$,
which differ by less than a few parts in 10$^3$.

In the non-condensed phase, $T\geq T_c$, the condensate density $n_0({\bf r})$ 
vanishes and the self-consistent procedure becomes easier because we do not
need to solve the equation (\ref{cequ1}) for the condensate. 

In the present work we have studied axially symmetric harmonic traps with the 
confining potential given by 
\begin{equation}
V_{ext}({\bf r}) = \frac{m}{2} (\omega_r^2\rho^2 + \omega_z^2 z^2) \;\;,
\label{Vext}
\end{equation}
where $\omega_r$ is the radial frequency in the $x-y$ plane. Clearly the 
calculation can be also applied to different confining potentials $V_{ext}$.
Before performing the calculation described above one must specify the trap 
parameters (frequency $\omega_r$ and deformation $\lambda=\omega_z/\omega_r$) 
as well as
the parameters describing the system (mass $m$ of the particles, scattering 
length 
$a$ and total number of atoms $N$). For a given temperature $T$ the 
above self-consistent procedure gives directly the chemical potential, $\mu$, 
the number of atoms in the condensate, $N_0(T)$, the condensate and 
non-condensate densities $n_0$ and $n_T$.

We have first checked the validity of the semiclassical approximation employed 
in the present work by comparing (see Fig. 2) our results for the condensate 
fraction with the ones recently obtained in Ref. \cite{HZG} by solving the full 
Bogoliubov equations (\ref{bogeqs}) at finite temperature. 
The various parameter characterizing the 
trap and the gas are the same in the two calculations. Despite the relatively
small value of $N$ (=2000) the agreement is good also at low $T$, revealing 
that no significant errors in the analysis are introduced by the use of the 
semiclassical approximation for the excited states.

\subsection{Condensate fraction and density distributions}

In Ref. \cite{GPS} we have already studied the effect of the interatomic 
potential 
on the temperature dependence of the condensate fraction and on the critical 
temperature. We found that a repulsive interaction among the atoms enhances 
the thermal depletion of the condensate with respect to the prediction of 
the non-interacting model for the same number of trapped atoms. The effect 
results in a shift of the critical temperature towards lower temperatures.
In Ref. \cite{GPS} we have also given an analytic expression for the shift 
$\delta T_c=T_c-T_c^0$ of the critical temperature from the prediction 
$k_BT_c^0=\hbar\omega(N/\zeta(3))
^{1/3}$ of the non-interacting model holding in the large $N$ limit. 
The result, valid to lowest order in the scattering length
and for large values of $N$, reads
\begin{equation}
\frac{\delta T_c}{T_c^0} \simeq -0.73\frac{\bar{\omega}}{\omega}N^{-1/3} 
-1.3 \frac{a}{a_{HO}}N^{1/6} \;\;.
\label{rdtc}
\end{equation}
The first term on the right hand side (r.h.s.) of the above equation  
is independent of the interaction and gives the first correction to $T_c^0$ 
due to the finite number of atoms in the 
trap (see discussion in section 2-C and \cite{IGAS}).
The second term on the r.h.s. of eq. (\ref{rdtc}) accounts for the effect
of interaction; it can be either positive or negative, depending on the sign of
$a$, increases with $N$ and depends only on the geometric mean $\omega$ 
through the harmonic oscillator length $a_{HO}=(\hbar/m\omega)^{1/2}$. 
In the currently available experimental situations the shift due to 
interaction is the dominant effect as soon as the number of particles is of 
the order of 10$^4$ for the JILA trap ($a/a_{HO}=7.35 \times 10^{-3}$, 
$\lambda=\sqrt{8}$) and of the order of 10$^5$ for the MIT trap 
($a/a_{HO}=2.55 \times 10^{-3}$, $\lambda=18/320$). The shift of the critical 
temperature due to interaction has a quite simple physical interpretation: the 
presence of repulsive interactions among the atoms acts in reducing the 
density of particles in the center of the trap. Thus lower temperatures are 
needed to reach there the critical density for 
Bose-Einstein condensation. The opposite happens in the case 
of attractive interactions. 

In Fig. 3 we present results for the temperature dependence of the condensate 
fraction obtained with our self-consistent mean-field method,
corresponding to $N=5 \times 10^4$ Rb atoms in the JILA trap, and to 
$N=2.9 \times 10^7$ Na atoms 
in the MIT trap. 
In the same figure we also plot the 
predictions for the same configurations obtained switching off the two-body 
interaction. The enhancement of the thermal depletion due to interaction 
is significant over the whole temperature range 
and the corresponding shift of the critical temperature is also
clearly visible, being in quantitative agreement with the analytic estimate 
(\ref{rdtc}). In Fig. 3 we also show the accuracy of the scaling behaviour
discussed in section 2-D. Both configurations correspond to the value 
$\eta=0.45$ of the scaling parameter. The solid curve is the scaling function 
obtained from    
(\ref{N0red}) with $\eta=0.45$, and the figure clearly shows that scaling is 
very well verified for these configurations and that hence extremely different 
experimental situations can give rise to the same thermodynamic behaviour. 
Only very close to $T_c$ the $N=5 \times 10^4$ points exhibit small deviations 
from the scaling behaviour. In fact close to $T_c$ the scaling law 
(\ref{N0red}) is approached more slowly with increasing $N$. It is interesting
to notice that finite size effects, responsible for the deviations from the 
scaling law $1-t^3$ in the non-interacting model (see eq. (\ref{NT01})), are
still visible for these configurations, whereas they are strongly quenched
in the presence of the interaction.

In Fig. 4 we present results for the condensate fraction $N_0/N$ as a function 
of the reduced temperature $t$ for three different values of the scaling 
parameter $\eta$, covering the presently available experimental conditions 
(see also Table I). In the figure
the open diamonds with the error bars are the result of the Monte-Carlo 
simulation of Ref. \cite{KRA} which correspond to the value $\eta=0.33$ and 
which are in good agreement with our predictions. The dots are the 
experimental results of Ref. \cite{JIL}. In the experiments the number of 
particles $N$ varies with $t$, with the value of $\eta$ ranging from 0.45 to
0.39. The experiments exhibit smaller deviations from the non-interacting 
curve with respect to the ones predicted by theory. One should however keep in 
mind that the measured value of $T$ is obtained from the velocity 
distribution of the thermal particles after expansion. 
The identification of this temperature with 
the one of the system before expansion ignores the interaction with the 
condensate which is expected to produce an acceleration of the thermal cloud.
A preliminary estimate based on the calculation of the interaction energy 
between the thermally excited atoms and the condensate, shows that 
for $T\simeq 0.5 T_c^0$ the final kinetic 
energy of the thermal cloud could be about 10\% larger than its value before 
expansion.

In Fig. 5 we show the temperature dependence of the density of particles in 
the center of the trap for two configurations: $N=10^7$ Na atoms in the 
MIT trap  
and 
$N=5000$ Rb atoms in the JILA trap. 
The BEC transition is clearly marked by the discontinous 
change of slope in the two curves.

In the limit of large numbers of particles the density in the center of the 
trap exhibits a scaling behaviour in terms of the parameters $t$ and $\eta$ 
when expressed in the dimensionless form discussed in section 2-D.  
In Fig. 6 the peak value of the reduced density $\tilde{n}(\tilde{{\bf r}}=0)=
n({\bf r}=0)(2k_BT_c^0/m\omega^2)^{3/2}/N$ is reported as a function of the 
reduced temperature $t$ for three values of the scaling parameter $\eta$
(see also Table II).
At $T=0$ one has 
$\tilde{n}(\tilde{{\bf r}}=0)=8\pi\eta^{-3/2}/15$ which  
decreases by increasing $\eta$. For $T\gg T_c^0$ the system 
approaches a classical ideal gas in the confining potential $V_{ext}$ 
where $\tilde{n}(\tilde{{\bf r}}=0)=(\pi t)^{-3/2}$. 
Notice that in the non-interacting model the peak density  
does not 
exhibit scaling behaviour for $T<T_c^0$.
For example at $T=0$ one has $\tilde{n}(\tilde
{{\bf r}}=0)=(2/\pi)^{3/2}(N/\zeta(3))^{1/2}$ which diverges as 
$N\rightarrow\infty$.   

In Figs. 7 a-d we show the density profiles of the total, condensate and 
non-condensate densities for four values of $T/T_c^0$ obtained from the 
self-consistent solution of the equations of the Popov approximation 
for the configuration
corresponding to $N=5000$ Rb atoms in the JILA trap. 
The density profiles are plotted as a function of the radial coordinate. 
Similar profiles are obtained for 
the $z$-direction.
At $T/T_c^0=0.3$ 
(7-a) the thermal density $n_T$ 
is very small compared to the condensate density  
which almost coincides with the total density. By increasing $T/T_c^0$ (7-b 
and 7-c) $n_T$ increases significantly and at $T/T_c^0=0.9$ (7-c) 
it becomes 
comparable with the condensate density in the center of the trap. 
From Figs. 7-b, 7-c
it clearly appears that the thermal density is not monotonically 
decreasing, and reaches a flat maximum near the boundary of the condensate.  
This behaviour arises from the repulsive 
interaction with the atoms in the condensate. Above $T_c^0$
(7-d) the condensate has disappeared and the thermal cloud has significantly
spread out. In Figs. 7 a-d the scale on the vertical axis has been kept 
the same to give a visible evidence of the decrease of the density in the 
central region of the trap.
The figure clearly shows the distinct behaviour  
of the two components of the gas: the condensate occupies the 
internal region of the trap while the thermal component forms a much broader 
cloud.  

In Fig. 8 we show the mean square radii of the condensate and 
of the thermal component as a function of $T/T_c^0$ for the same configuration 
of Fig. 7. The short-dashed line refers to 
the mean square radius in the 
$x$-direction of the condensate atoms. By increasing $T$ it 
decreases slowly and approaches a finite value at the transition point. The 
average radius of the thermal cloud
is represented by the long-dashed 
line. It is always larger than the condensate radius and increases almost 
linearly 
with temperature. The solid line corresponds to average radius of the total 
cloud
$\int d^3{\bf r} n({\bf r})x^2/N$ which coincides with
the condensate radius at $T=0$ and goes over to the non-condensate radius for
$T>T_c$. 

In the limit of large $N$ the radii of the atomic cloud exhibit a 
scaling behaviour when expressed in the reduced units of section 2-D
($\tilde{r}_i=\sqrt{m/2k_BT_c^0}\omega_ir_i$). In Figs. 9-10 we show how the
scaling behaviour is approached, along the $x$  
($\sqrt{\langle\tilde{x}^2\rangle}$) and $z$-direction 
($\sqrt{\langle\tilde{z}^2\rangle}$) respectively, for the JILA (open circles) 
and the MIT (solid circles) configuartions. Fig. 9 shows that  
the scaling behaviour for
$\sqrt{\langle\tilde{x}^2\rangle}$ 
is well verified for both configurations, whereas 
along the $z$-direction the configuration with 
the smallest number of particles ($N=5\times 10^4$) 
exhibit a slight deviation from the scaling behaviour (see Fig. 10). 
This reveals that in the JILA trap  
finite size 
effects along the direction of stronger confinement have not completely
disappeared for this value of $N$. In the same figures we also plot the result 
for the non-interacting model ($\eta=0$). The dotted line is the result in 
the large $N$ scaling limit, while the open and solid diamonds take into 
account finite size effects. 
The deviations from the scaling law
are much more evident in the non-interacting case. 

In Fig. 11 we plot the reduced mean radius of the cloud $\sqrt{\langle\tilde
{r}_i^2\rangle}$ as a function of the reduced temperature $t$ for three values
of the scaling parameter $\eta$ (see also Table III). 
The increase of the radius due to 
interaction is clearly visible by comparing the three curves with the 
result holding in the non-interacting model for $t<1$: 
$\sqrt{\langle\tilde{r}_i^2
\rangle}=
\sqrt{E/6Nk_BT_c^0}=\sqrt{\zeta(4)/2\zeta(3)}t^2$ (see eq. (\ref{E001})).  
Notice that for spherical traps the mean square radius 
is proportional to the  
harmonic potential energy $\langle\tilde{r}_i^2\rangle=E_{HO}/3Nk_BT_c^0$.   
        
\subsection{Energy and specific heat}

In this section we discuss the energetics of the trapped gases by presenting
results for the chemical potential, total energy, release energy 
and specific heat. For all these quantities the scaling 
behaviour is well verified in the configurations realized in the experiments 
and 
we will therefore discuss our results only in the scaling limit 
(see section 2-D).

In Fig. 12 we report the results for the chemical potential in units of 
$k_BT_c^0$ as a function of the reduced temperature $t$ for three values of 
the scaling parameter $\eta$ (see also Table IV). Notice that for 
$t\rightarrow 0$ the plotted quantity coincides with $\eta$ (see definition 
(\ref{ratio1})). 
In the classical limit, $T\gg T_c^0$, the dependence on the 
interaction parameter disappears and one finds the classical ideal gas 
prediction $\mu/k_BT_c^0=t\log(\zeta(3)/t^3)$.

In Fig. 13 we present the results for the total 
energy $E$ of the system (see also Table V). 
At high temperature the behaviour is given by the 
classical law $E/Nk_BT_c^0 = 3t$. In the BEC phase the energy per particle 
is significantly larger than the prediction of the non-interacting model. 
This feature has been confirmed by the first experimental results on the 
temperature dependence of the release energy, which is the sum of the 
kinetic and interaction energy \cite{JIL}. In Fig. 13 the theoretical 
predictions for the release energy per particle, in units of $k_BT_c^0$, are 
shown together with the experimental results of Ref. \cite{JIL} (see also 
Table VI). Though the experimental points below $T_c$ lie well above the 
non-interacting curve giving evidence for interaction effects, the accuracy 
is not enough to make a quantitative comparison with theory.  

Finally in Fig. 15 we plot the specific heat per particle as a function of the 
reduced temperature $t$. These curves coincide with the derivative 
with respect 
to $t$ of the curves of Fig. 13. The effect of interaction
is to round off the peak at the transition with respect to the 
non-interacting model. Direct experimental 
results on the specific heat are not yet available. An indirect estimate 
can be obtained by differentiating the experimental curve  
of the release energy (see Ref. \cite{JIL}).
One should however keep in mind that the release energy is roughly a factor 
two smaller than the total energy. 
 
\subsection{Moment of inertia}

The superfluid properties of the trapped gases can be investigated by 
calculating the ratio between the moment of inertia of the system $\Theta$ 
and its rigid value $\Theta_{rig}$ (see section 2-A). In Fig. 16 our 
predictions for $\Theta/\Theta_{rig}$ are shown as a function of the reduced
temperature $t$. We find that, differently from the other quantities 
discussed above, this ratio does not exhibit a significant dependence on 
the scaling parameter $\eta$. The figure also shows the very different 
behaviour 
exhibited by the non-interacting model where, even for large values 
of $N$, there is a strong dependence on the number of trapped atoms and 
the anisotropy $\lambda$ of the trap (see eq. (\ref{mi0})).

\section{Comparison with the homogeneous Bose gas}

The thermodynamic behaviour of the trapped Bose gas discussed in the previous 
sections exhibits important differences with respect to the case of the uniform
gas. The purpose of this section is to discuss these differences and to 
explain their physical origin. We will mainly focus on the temperature 
dependence of the condensate fraction where the effects are particularly 
visible and significant.

Let us first discuss the behaviour of the ideal non-interacting gas. In the 
uniform case the critical temperature for the onset of Bose-Einstein 
condensation is given by the well known expression
\begin{equation}
k_BT_c^0 = \frac{2\pi\hbar^2}{m}\left(\frac{n}{\zeta(3/2)}\right)^{2/3} 
\label{utc}
\end{equation} 
as a function of the density $n$ of the gas. In the presence of harmonic 
confinement the critical temperature instead depends on the number of atoms 
and on the oscillator frequency according to the expression (\ref{Tc})
\begin{equation}
k_BT_c^0 = \hbar\omega \left(\frac{N}{\zeta(3)}\right)^{1/3} \;\;,
\label{utc1}
\end{equation}
holding in the large $N$ limit. By introducing the reduced temperature 
$t=T/T_c^0$ one can naturally compare the thermodynamic behaviour of the two
systems. In particular the temperature dependence of the condensate fraction
follows the laws
\begin{equation}
\frac{N_0}{N} = 1 - t^{3/2} \;\;,
\label{ucond}
\end{equation}
and
\begin{equation}
\frac{N_0}{N} = 1 - t^3 \;\;,
\label{ucond1}
\end{equation}
in the uniform and in the harmonically trapped gases respectively. Different 
laws for the temperature dependence of the condensate are predicted to 
occur by changing the form of the trapping potential \cite{BPK}.

A major difference between the two systems concerns the relative shape of the 
condensate and of the thermally excited densities (see Fig. 7). In fact, 
differently from the uniform gas where the two components obvously overlap 
everywhere, in the presence of the external harmonic potential the size of 
the thermal component is larger than the one of the condensate. The former is 
in fact of the order of the classical thermal radius $(2k_BT/m\omega^2)^{1/2}$,
while the latter is fixed by the harmonic oscillator length $a_{HO}=(\hbar/m
\omega)^{1/2}$. As a consequence, as clearly shown by Fig. 7, most of the atoms 
of the thermal component lie outside the condensate and form a much more dilute
gas, practically for any value of $T$. This 
behaviour has important consequences when we switch on the two-body 
interaction. The effect of the repulsive interaction is in fact twofold. One
the one hand it produces an expansion of the condensate whose
size can significantly increase with respect to the one of the 
non-interacting model (this effect is fixed by the dimensionless 
parameter $Na/a_{HO}$ as discussed in Refs. \cite{OX1,BP,DS}). On the other
hand at finite temperature the repulsive forces favour the depletion of the 
condensate. In fact atoms are energetically favoured to leave the 
condensate where the repulsion is strong, and to reach the non-condensate 
phase which is much more dilute and interaction effects are consequently less 
important. This is exactly the opposite of what happens in the homogenous   
gas where the repulsion effects are stronger in the thermal component 
than in the condensate. This different behaviour can be understood by 
considering, for example, the Hartree-Fock expression for the interaction 
energy (last term of eq. (\ref{hfen}), or, more directly, the Hartree-Fock
expression for the particle distribution function  
(we use here 
the Hartree-Fock formalism because the effect can be discussed in a 
particularly transparent way in this approximation; analogous results are 
obtained in the Popov approximation as proven by the numerical results 
reported in Fig. 17 (a)),
\begin{equation}
F({\bf p},{\bf r}) = \left( \exp((p^2/2m+V_{ext}-\mu+2gn)/k_BT)-1 \right)
^{-1} \;\;,
\label{upd}
\end{equation}
which is obtained from eqs. (\ref{pdhf}) and (\ref{epshf}). 
In the homogeneous case one has 
$V_{ext}=0$, $\mu=2gn-gn_0$ and the interaction effects produce
a {\it quenching} of $F$ and hence of the thermal depletion $N_T=\int 
d{\bf p}d{\bf r}/(2\pi\hbar)^3 F({\bf p},{\bf r})$ with respect to the ideal 
gas. In the case of harmonic confinement one sees that the relevant region 
of space contributing to $N_T$ lies outside the condensate where the term 
$2gn$ is negligible and interaction effects enter only through the chemical 
potential. In this case the repulsive interactions produce 
an {\it enhancement} of 
$F$ with respect to the ideal gas, since the value of $\mu$ is positive for
$T<T_c$. Notice that the value of the chemical potential is strongly affected 
by the interaction effects in the condensate and consequently the final 
effects on the thermal depletion can be sizable. In Fig. 17 (a-b) we show 
explicitly the temperature dependence of the condensate fraction in the 
uniform (a) and in the trapped (b) gases. In order to make the comparison 
significant, the density of the uniform gas 
has been fixed equal to the peak density $n({\bf r}=0)$ of the trapped gas 
at zero temperature. The figure clearly shows that the
effect of the interaction goes in the opposite direction in the two cases.

An important consequence of the above results is that in a trapped Bose gas the
repulsive interactions have the effect of {\it lowering} the critical 
temperature. This can be also understood by noting that the interaction 
produces a general expansion of the system and one has finally to deal with 
a more dilute gas with respect to the non-interacting gas with the same value
of $N$. In the uniform case, where the density is kept fixed, interaction 
effects are known to produce instead an {\it increase} of $T_c$, at least for 
very dilute gases. The corresponding shift cannot however be calculated in 
mean field theory, being associated with many-body fluctuation effects typical 
of critical phenomena. Also the behaviour of the condensate fraction near the 
critical point looks different in the two cases (see Fig. 17). In the uniform 
gas the mean field approach gives rise to a sizable gap in the condensate 
fraction at $T_c$. This gap, which is of the order of $(a^3n)^{1/3}$, is an 
artifact of the theory and vanishes when 
fluctuation effects are 
taken into account near the critical point \cite{STOOF}. 
This gap is present also in the case of the trapped gas, though it has a very 
small   
effect on the condensate fraction and on other thermodynamic functions.
The reason is that the size of the condensate cloud near the transition 
point is small and the gap in the condensate fraction turns out to be 
of the order of $\eta^{10}$ \cite{N4}.

Figure 17 points out also another important difference between the uniform and 
trapped gases. In the latter case phonons 
are always important in the calculation of the condensate fraction (this is 
proven by the significant difference between the prediction of the 
Hartree-Fock theory and the full Popov 
calculation which includes phonon effects). On the other hand
in the case of harmonic 
confinement the Hartree-Fock approximation turns out to be an excellent 
approximation also at relatively low temperatures. In fact for these 
temperatures most of the 
contribution to 
thermodynamic properties of the trapped gas arise from excitations lying  
close to the classical boundary where  
the Hartree-Fock theory is a good 
approximation. 

It is finally worth noting that the ratio between the $T=0$ value of the
chemical potential and the critical temperature depends on the gas parameter 
$a^3n$ quite differently in the two cases. In fact while in the trapped gas 
this ratio behaves as $(a^3n({\bf r}=0))^{1/6}$ (see eq. (\ref{ratio1})), 
in the uniform gas the dependence is much stronger and is given by 
$(a^3n)^{1/3}$.

In conclusion we have shown that a Bose gas in harmonic traps exhibits a 
thermodynamic behaviour quite different from the one of the uniform gas. 
The difference mainly originates from the different structure of the 
elementary excitations and from their coupling with the condensate. This should
be taken into account when one compares the predictions of many-body theories 
obtained in uniform gases with the properties of these novel trapped systems.

\section{Conclusions}

In this paper we have presented a systematic discussion of the thermodynamic 
behaviour of an interacting Bose gas confined in a harmonic trap. The formalism
was based on an extension of Bogoliubov theory to non uniform systems at 
finite temperature. By using the semiclassical WKB approximation for the 
excited states we have been able to investigate various thermodynamic 
functions for a wide range of the relevant parameters of the system. The main 
results emerging from our analysis are briefly summarized below:

i) For large numbers of atoms in the trap the effects of two-body interactions 
(limited in this work to the repulsive case) can be accounted for by the 
scaling parameter
\begin{equation}
\eta = {\mu_0 \over k_BT_c^0} = 1.57 (N^{1/6}
{a\over a_{HO}})^{2/5} \;\;,
\label{ratio2}
\end{equation}
which gives the ratio between the $T=0$ value of the chemical potential and of 
the critical temperature for Bose-Einstein condensation. Typical values of 
$\eta$ in the available experiments range between 0.38 to 0.45. Physical 
systems with very different values of $N$, scattering length and oscillator 
frequencies are predicted to give rise to the same thermodynamic behaviour 
provided they correspond to the same value of $\eta$.  

ii) Interactions provide a significant quenching of the condensate fraction 
with respect to the non-interacting model and a shift of the critical 
temperature towards lower values. This behaviour, which differs from the one 
exhibited by a uniform Bose gas, points out a unique and interesting feature
exhibited by these trapped systems: the occurrence of elementary excitations 
located outside the condensate. The thermal excitation of these modes is 
energetically favoured by the presence of the repulsive interaction and plays 
a crucial role in the thermodynamics of the system.

iii) Interactions are also responsible for a significant enhancement of the 
thermal excitation energy. Evidence for such an effect has been recently 
become available from first measurements of the release energy. 

iv) Results have been also obtained for the moment of inertia whose deviations 
from the rigid value point out the occurrence of superfluidity. Such 
deviations are significant in a rather wide range of temperatures below $T_c$ 
and have been found to depend very little on the value of the interaction 
parameter $\eta$.

v) Explicit comparison with the predictions of Path Integral Monte Carlo 
simulations as well as 
with the full solutions of the discretized Bogoliubov equations (without 
using the WKB approximation) have revealed that the method presented in this 
paper is remarkably accurate in the relevant range of temperatures. 
Experimental results instead are not yet enough accurate to check the 
theoretical predictions in a quantitative way.

\section*{Acknoledgments}

We gratefully acknowledge useful discussions with Franco Dalfovo and Allan 
Griffin. We would also like to thank E. Zaremba for providing the numerical 
results of Ref. \cite{HZG}.

\begin{figure}
\caption{Temperature dependence of the condensate fraction in the 
non-interacting limit. The solid circles correspond to the exact quantum 
calculation for $N=1000$ atoms in the JILA trap and the solid line to the 
semiclassical approximation of eq. (\ref{NT01}). The short-dashed line
refers to the large $N$ limit.}
\end{figure}

\begin{figure}
\caption{Condensate fraction as a function of $T/T_c^0$. The solid circles 
are the result of the calculation of Ref. \protect\cite{HZG} for $N=2000$ 
Rb atoms 
in an isotropic trap with oscillator length $a_{HO}=7.62\times 10^{-5}$ cm,
obtained without using the semiclassical approximation. The solid line is 
the result for the same configuration employing the semiclassical
approximation. Also plotted is the curve of the non-interacting model in the 
large $N$ limit (short-dashed line).} 
\end{figure}

\begin{figure}
\caption{Theoretical predictions for the condensate fraction as a function 
of $T/T_c^0$.  
The open circles correspond to $N=5\times 10^4$ Rb atoms in the JILA 
trap ($a/a_{HO}=7.35\times 10^{-3}$, $\lambda=\protect\sqrt{8}$). 
The solid circles correspond to $N=2.9\times 10^7$ Na atoms in the 
MIT trap ($a/a_{HO}=2.55\times 10^{-3}$, $\lambda=18/320$). 
The solid line corresponds to the scaling limit with $\eta=0.45$.
The dotted line is the $1-t^3$ curve of the non-interacting 
model in the large $N$ limit. The open and solid diamonds correspond to 
$N=5\times 10^4$ and $N=2.9\times 10^7$ non-interacting particles in the JILA 
and MIT traps respectively.}
\end{figure}

\begin{figure}
\caption{Condensate fraction in the scaling limit as a function of $T/T_c^0$ 
for three values of 
$\eta$. Solid line: $\eta=0.45$, long-dashed line: 
$\eta=0.39$, short-dashed line: $\eta=0.31$. Open diamonds: PIMC results 
of Ref. \protect\cite{KRA}. Solid circles: experimental results from Ref. 
\protect\cite{JIL}.
The dotted line refers to the non-interacting model in the large $N$ limit.}
\end{figure}

\begin{figure}
\caption{Peak density as a function of $T/T_c^0$. Solid line: $N=10^7$ Na 
atoms in the MIT trap, long-dashed line: $N=5\times 10^3$ Rb atoms in 
the JILA trap.}
\end{figure}
 
\begin{figure}
\caption{Reduced peak density as a function of $T/T_c^0$ for three values of 
the scaling parameter $\eta$ (see Fig. 4).}
\end{figure}

\begin{figure}
\caption{Density profiles as a function of the radial coordinate at $z=0$ 
for $N=5\times 10^3$ Rb 
atoms in the JILA trap 
at different temperatures. Solid line: total density, long-dashed line: 
condensate density, dotted line: non-condensate density. Lengths are in 
units of the radial oscillator length $a_r=(\hbar/m\omega_r)^{1/2}$. Densities 
are in units of $a_r^{-3}$.}
\end{figure}

\begin{figure}
\caption{Root mean square (r.m.s.) radii, in units of $a_r$, 
in the $x$-direction for 
$N=5\times 10^3$ Rb atoms 
in the JILA trap as a function of $T/T_c^0$. Solid line: average radius 
of the total cloud; long-dashed line: average radius of the thermal cloud;
short-dashed line: average radius of the condensate cloud.}
\end {figure}

\begin{figure}
\caption{ r.m.s. radii in the $x$-direction in reduced units as 
a function of $T/T_c^0$ for $\eta=0.45$ (solid line).  
The open circles refer to $N=5\times 10^4$ Rb atoms in the JILA 
trap. The solid circles correspond to $N=2.9\times 10^7$ Na atoms in the 
MIT trap. The dotted line refers to the non-interacting 
model in the large $N$ limit. The open and solid diamonds correspond to 
$N=5\times 10^4$ and $N=2.9\times 10^7$ non-interacting particles in the JILA 
and MIT traps respectively.}
\end{figure}

\begin{figure}
\caption{r.m.s. radii in the $z$-direction (see Fig. 9).} 
\end{figure}

\begin{figure}
\caption{r.m.s. radii in reduced units as a function of 
$T/T_c^0$ for three values of 
the scaling parameter $\eta$ (see Fig. 4). The dotted line refers to the 
non-interacting model in the large $N$ limit.}
\end{figure}

\begin{figure}
\caption{Chemical potential as a function of $T/T_c^0$ for three values of 
the scaling parameter $\eta$ (see Fig. 4). The dotted line refers to the 
non-interacting model in the large $N$ limit.}
\end{figure}

\begin{figure}
\caption{Total energy of the system as a function of $T/T_c^0$ for three 
values of the scaling parameter $\eta$ (see Fig. 4). The dotted line refers 
to the non-interacting model in the large $N$ limit.}
\end{figure}
  
\begin{figure}
\caption{Release energy of the system as a function of $T/T_c^0$ for three 
values of the scaling parameter $\eta$ (see Fig. 4). 
The solid circles are the experimental results of Ref. \protect\cite{JIL}.
The dotted line refers 
to the non-interacting model in the large $N$ limit.}
\end{figure}

\begin{figure}
\caption{Specific heat as a function of $T/T_c^0$ for three 
values of the scaling parameter $\eta$ (see Fig. 4). The dotted line refers 
to the non-interacting model in the large $N$ limit.}
\end{figure}

\begin{figure}
\caption{Ratio $\Theta/\Theta_{rigid}$ as a function of $T/T_c^0$ for three 
values of the scaling parameter $\eta$ (see Fig. 4). The three curves 
coincide almost exactly and are represented by the solid line. The dotted line 
refers to $N=5\times 10^4$ atoms in the JILA trap in the non-interacting 
model. The dot-dashed line refers to $N=2.9\times 10^7$ atoms in the MIT 
trap in the non-interacting model.}
\end{figure}

\begin{figure}
\caption{Condensate fraction as a function of $T/T_c^0$. (a) refers to a 
homogeneous gas with the gas parameter $a^3n=6.2\times 10^{-5}$, (b) refers to
a trapped gas with the same value of the gas parameter in the center of the 
trap at $T=0$, corresponding to the value $\eta=0.45$.  
Solid line: Popov approximation; long-dashed line: Hartree-Fock approximation; 
dotted line: $1-t^{3/2}$ (a) and $1-t^3$ (b) curves of the non-interacting 
model in the homogeneous and trapped case respectively.}
\end{figure}

\begin{table}
\caption{Condensate fraction}
\begin{tabular}{cccc}
$T/T_c^0$ & $N_0/N$     & $N_0/N$     & $N_0/N$     \\
$\;$      & $\eta=0.31$ & $\eta=0.39$ & $\eta=0.45$ \\
\tableline
0.09      & 0.99        & 0.99        & 0.99        \\
0.25      & 0.95        & 0.94        & 0.94        \\
0.50      & 0.76        & 0.73        & 0.71        \\
0.75      & 0.38        & 0.33        & 0.29        \\
0.91      & 0.08        & 0.05        & 0.03        \\
\end{tabular}  
\end{table}

\begin{table}
\caption{Peak density in reduced units}
\begin{tabular}{cccc}
$T/T_c^0$ & $\tilde{n}(\tilde{{\bf r}}=0)$ & $\tilde{n}(\tilde{{\bf r}}=0)$
& $\tilde{n}(\tilde{{\bf r}}=0)$     \\
$\;$      & $\eta=0.31$ & $\eta=0.39$ & $\eta=0.45$ \\
\tableline
0.25      & 3.40        & 2.43        & 1.95        \\
0.50      & 3.19        & 2.28        & 1.81        \\
0.75      & 2.55        & 1.79        & 1.41        \\
1.00      & 0.30        & 0.28        & 0.26        \\
1.25      & 0.15        & 0.15        & 0.15        \\
\end{tabular}  
\end{table}

\begin{table}
\caption{Root mean square radius in reduced units}
\begin{tabular}{cccc}
$T/T_c^0$ & $\protect\sqrt{\langle\tilde{r}_i^2\rangle}$ & $\protect
\sqrt{\langle\tilde{r}_i^2\rangle}$ & $\protect\sqrt{\langle\tilde{r}_i^2
\rangle}$ \\
$\;$      & $\eta=0.31$ & $\eta=0.39$ & $\eta=0.45$ \\
\tableline
0.25      & 0.22        & 0.24        & 0.26        \\
0.50      & 0.29        & 0.31        & 0.33        \\
0.75      & 0.46        & 0.48        & 0.50        \\
1.00      & 0.67        & 0.68        & 0.68        \\
1.25      & 0.77        & 0.78        & 0.78        \\
\end{tabular}  
\end{table}

\begin{table}
\caption{Chemical potential}
\begin{tabular}{cccc}
$T/T_c^0$ & $\mu/k_BT_c^0$ & $\mu/k_BT_c^0$
& $\mu/k_BT_c^0$     \\
$\;$      & $\eta=0.31$ & $\eta=0.39$ & $\eta=0.45$ \\
\tableline
0.25      &  0.31       &  0.39       &  0.45       \\
0.50      &  0.30       &  0.37       &  0.43       \\
0.75      &  0.24       &  0.31       &  0.36       \\
1.00      &  0.02       &  0.03       &  0.04       \\
1.25      & -0.70       & -0.69       & -0.69       \\
\end{tabular}  
\end{table}

\begin{table}
\caption{Total energy}
\begin{tabular}{cccc}
$T/T_c^0$ & $E/Nk_BT_c^0$ & $E/Nk_BT_c^0$
& $E/Nk_BT_c^0$     \\
$\;$      & $\eta=0.31$ & $\eta=0.39$ & $\eta=0.45$ \\
\tableline
0.25      & 0.24        & 0.30        & 0.35        \\
0.50      & 0.47        & 0.55        & 0.61        \\
0.75      & 1.27        & 1.38        & 1.47        \\
1.00      & 2.73        & 2.75        & 2.77        \\
1.25      & 5.59        & 3.61        & 3.62        \\
\end{tabular}  
\end{table}

\begin{table}
\caption{Release energy}
\begin{tabular}{cccc}
$T/T_c^0$ & $E_R/Nk_BT_c^0$ & $E_R/Nk_BT_c^0$
& $E_R/Nk_BT_c^0$     \\
$\;$      & $\eta=0.31$ & $\eta=0.39$ & $\eta=0.45$ \\
\tableline
0.25      & 0.10        & 0.12        & 0.14        \\
0.50      & 0.22        & 0.25        & 0.28        \\
0.75      & 0.63        & 0.68        & 0.72        \\
1.00      & 1.36        & 1.37        & 1.38        \\
1.25      & 1.79        & 1.80        & 1.81        \\
\end{tabular}  
\end{table}


\begin{references}
\bibitem{N1} Corrections to (\ref{Tc}) due to finite size effects have been the
object of several recent papers \cite{IGAS}.
\bibitem{N3} In the limiting case $\eta\rightarrow 0$ the equations can be
solved analytically. For example for the condensate fraction (see eq. 
(\ref{N0red})) one finds the expansion:
$N_0/N=(1-t^3) - \zeta(2)\eta t^2(1-t^3)^{2/5} /\zeta(3)$.
\end{references}
\end{document}